\documentclass[aip,jcp,preprint,groupedaddress]{revtex4-1}%
\usepackage{graphicx}
\usepackage{tabularx}
\usepackage{dcolumn}
\usepackage{longtable}
\usepackage{tensor}
\usepackage{color}
\usepackage{placeins}
\usepackage{bm}
\usepackage{amsmath}
\usepackage{amsfonts}
\usepackage{amssymb}%
\providecommand{\U}[1]{\protect\rule{.1in}{.1in}}
\providecommand{\U}[1]{\protect\rule{.1in}{.1in}}
\begin{document}
\title{Efficient geometric integrators for nonadiabatic quantum dynamics. I. The
adiabatic representation}
\author{Seonghoon Choi}
\author{Ji\v{r}\'{\i} Van\'{\i}\v{c}ek}
\email{jiri.vanicek@epfl.ch}
\affiliation{Laboratory of Theoretical Physical Chemistry, Institut des Sciences et
Ing\'enierie Chimiques, Ecole Polytechnique F\'ed\'erale de Lausanne (EPFL),
CH-1015, Lausanne, Switzerland}
\date{\today}

\begin{abstract}
Geometric integrators of the Schr\"{o}dinger equation conserve exactly
many invariants of the exact solution. Among these integrators, the
split-operator algorithm is explicit and easy to implement, but,
unfortunately, is restricted to systems whose Hamiltonian is separable into a
kinetic and potential terms. Here, we describe several implicit
geometric integrators applicable to both separable and non-separable
Hamiltonians, and, in particular, to the nonadiabatic molecular Hamiltonian in
the adiabatic representation. These integrators combine the dynamic Fourier
method with recursive symmetric composition of the trapezoidal rule or
implicit midpoint method, which results in an arbitrary order of accuracy in
the time step. Moreover, these integrators are exactly unitary, symplectic,
symmetric, time-reversible, and stable, and, in contrast to the
split-operator algorithm, conserve energy exactly, regardless of the accuracy
of the solution. The order of convergence and conservation of
geometric properties are proven analytically and demonstrated numerically on a
two-surface NaI model in the adiabatic representation. Although each
step of the higher order integrators is more costly, these algorithms become
the most efficient ones if higher accuracy is desired; a thousand-fold speedup
compared to the second-order trapezoidal rule (the Crank-Nicolson method) was
observed for wavefunction convergence error of $10^{-10}$. In
a companion paper [J. Roulet, S. Choi, and J. Van\'{\i}\v{c}ek
(2019)], we discuss analogous, arbitrary-order compositions of the
split-operator algorithm and apply both types of geometric integrators to a
higher-dimensional system in the diabatic representation.

\end{abstract}
\maketitle

\graphicspath{{./figures_new/}{C:/Users/Jiri/Dropbox/Papers/Chemistry_papers/2018/Integrators_adiabatic/figures_new/}{"d:/Group Vanicek/Desktop/Choi/Integrators_adiabatic/figures_new/"}}

\section{\label{sec:intro}Introduction}

Separating
electronic from nuclear degrees of freedom leads to the Born--Oppenheimer
approximation\cite{Born_Oppenheimer:1927,book_Heller:2018} and the intuitive
picture of electronic potential energy surfaces. However, many chemical,
physical, and biological processes can only be described by taking into
account the correlation between the nuclear and electronic
motions,\cite{Mokhtari_Zewail:1990} which is reflected in the nonadiabatic
couplings between different Born--Oppenheimer
surfaces.\cite{book_Baer:2006,Domcke_Yarkony:2012,book_Nakamura:2012,book_Takatsuka:2015,Bircher_Rothlisberger:2017}
To address such processes, one can forget the Born--Oppenheimer picture and
treat electrons and nuclei on the same
footing,\cite{Shin_Metiu:1995,Albert_Engel:2016} use an exact
factorization\cite{Abedi_Gross:2010,Cederbaum:2008} of the molecular
wavefunction, or, most commonly, determine which Born--Oppenheimer states are
significantly coupled\cite{Zimmermann_Vanicek:2010,Zimmermann_Vanicek:2012}
and then solve the time-dependent Schr\"{o}dinger equation with a molecular
Hamiltonian that contains the nonadiabatic couplings. Below, we will only
consider the third, yet the most traditional way to treat quantum nonadiabatic dynamics.

An approach particularly suited to study the nonadiabatic population dynamics
of large chemical systems is the \textit{ab initio} multiple
spawning\cite{Ben-Nun_Martinez:2000, Curchod_Martinez:2018} and related
methods, all of which represent the wavefunction by a superposition
of time-dependent Gaussian basis functions moving along
classical\cite{Shalashilin_Child:2001a,Makhov_Shalashilin:2017} or
variational\cite{Worth_Burghardt:2004, Richings_Lasorne:2015} trajectories. If
high accuracy is required and especially if the Hamiltonian can be expressed
as a sum of products of one-dimensional operators, a nonadiabatic algorithm of
choice is the multiconfigurational time-dependent Hartree (MCTDH)
method\cite{Meyer_Cederbaum:1990, Worth_Burghardt:2008} or its multilayer
extension,\cite{Wang_Thoss:2003} which expand the state using orthogonal
time-dependent basis functions. The power of the MCTDH\ method relies on the
fact that only a small fraction of the tensor-product Hilbert space is
typically accessible during the time of interest; sparse-grid
methods\cite{Avila_Carrington:2017,book_Lubich:2008} also take advantage of
this phenomenon. However, there are systems, in which the full Hilbert space
is accessible, and then full grid or time-independent basis sets are
preferable.\cite{Kosloff:1988,book_Lubich:2008}

There also exist situations, where, in addition to prescribed accuracy, it
pays off to conserve certain invariants of the exact solution exactly,
regardless of the accuracy of the wavefunction. Because the
above-mentioned methods typically conserve none or only some of these
invariants, other methods, called geometric
integrators,\cite{book_Hairer_Wanner:2006} are needed in this setting. The
geometric integrators acknowledge that the Schr\"{o}dinger equation is
special, and not just another general differential equation. Using these
integrators can be likened to realizing that the Earth is not flat but round,
and even approximate models of its surface should take this curvature into
account. Geometric integrators are highly exploited in classical molecular
dynamics, where the deceptively simple Verlet algorithm,\cite{Verlet:1967,
Frenkel_Smit:2002} despite its only second-order accuracy, results in
exact conservation of $D$ invariants in a $D$-dimensional system, where $D$
can easily reach thousands or millions in state-of-the art simulations of proteins.

Time propagation schemes based on geometric integrators have also been applied
to the time-dependent Schr\"{o}dinger
equation.\cite{McCullough_Wyatt:1971,Feit_Steiger:1982, book_Lubich:2008,
book_Tannor:2007} Symmetric compositions of the first-order split-operator
algorithms,\cite{book_Lubich:2008, book_Tannor:2007} including the standard
second-order splitting,\cite{Feit_Steiger:1982} are unitary, symplectic,
stable, symmetric, and time-reversible regardless of the size of the time
step. Moreover, the symmetric split-operator algorithms can be recursively
composed to obtain efficient methods of arbitrary order in the time
step.\cite{Yoshida:1990, McLachlan:1995, Suzuki:1990,
book_Hairer_Wanner:2006,Wehrle_Vanicek:2011, Roulet_Vanicek:2019} In a
companion paper\cite{Roulet_Vanicek:2019} (below referred to as Paper II), we
implement such higher-order compositions for the nonadiabatic quantum
molecular dynamics in the diabatic representation.

Although the split-operator algorithms preserve numerous geometric properties
of interest of the exact evolution operator, their use is limited to systems
with Hamiltonians separable into a sum $\hat{H}=T(\hat{p})+V(\hat
{q})$ of two terms, the first depending only on the momentum
operator and the second only on the position operator. One must use a
different time propagation scheme for systems with a non-separable
Hamiltonian; for example, the nonadiabatic dynamics in the adiabatic
representation or particles in crossed electric and magnetic fields.

The explicit Euler method is the simplest integrator applicable to
non-separable Hamiltonians; it is, however,
unstable.\cite{book_Leimkuhler_Reich:2004, book_Hairer_Wanner:2006} The
implicit Euler method is stable regardless of the size of the time step but
requires solving a large, although sparse, system of linear equations at every
time step; furthermore, the method fails to preserve the unitarity, time
reversibility, energy conservation, and other geometric properties of the
exact evolution operator. The second-order differencing
method\cite{Askar_Cakmak:1978, Kosloff_Kosloff:1983, Leforestier_Kosloff:1991}
introduces symmetry by combining the forward and backward step of the explicit
Euler method. It is explicit and stable for small enough time steps, but does
not conserve the norm or energy exactly.

Another issue with the second-order differencing is that a much too small time
step is required to obtain an accurate solution.\cite{chapter_Lubich:2002}
This problem has been addressed by using the
Chebyshev\cite{Tal-Ezer_Kosloff:1984} and short iterative Lanczos
algorithms;\cite{Lanczos:1950,Park_Light:1986,Leforestier_Kosloff:1991} both
methods increase remarkably the efficiency of numerical integration by
effectively approximating the exact evolution operator. However, these two
methods are neither time-reversible nor symplectic, and the Chebyshev
propagation scheme does not even conserve the norm.

To address either the low accuracy or nonconservation of geometric properties
by various nonadiabatic integrators, we propose time propagation schemes based
on symmetric compositions of the trapezoidal rule (also known as the
Crank-Nicolson method\cite{Crank_Nicolson:1947,McCullough_Wyatt:1971}) or
implicit midpoint method. As we show below, because these elementary methods
are unitary, symplectic, energy conserving, stable, symmetric, and
time-reversible, so are their symmetric compositions. Furthermore,
like any other symmetric second-order algorithm, the trapezoidal rule and
implicit midpoint method can be recursively composed to obtain integrators of
arbitrary order of accuracy in the time step.\cite{Yoshida:1990,
McLachlan:1995, Suzuki:1990, book_Hairer_Wanner:2006} Methods with higher
orders of accuracy are useful for obtaining highly accurate solutions because,
for that purpose, they are more efficient than the second-order algorithms.
Although each time step of a higher-order method costs more, the solution with
the same accuracy can be obtained using a larger time step and, hence, a
smaller total number of steps in comparison to lower-order methods. The final
benefit of the proposed geometric integrators is the simple, abstract, and
general implementation of the compositions of the trapezoidal rule and
implicit midpoint methods; indeed, even these \textquotedblleft
elementary\textquotedblright\ methods are, themselves, compositions of simpler
explicit and implicit Euler methods.

In the adiabatic representation, the proposed integrators cannot be
fully compared with the integrators based on the compositions of the
split-operator algorithm, which are only applicable to separable Hamiltonians.
Both types of integrators, however, can be used in the diabatic
representation, which is the focus of Paper II.\cite{Roulet_Vanicek:2019} We,
therefore, compare the two integrators there, using a one-dimensional
model\cite{Engel_Metiu:1989} of NaI and a three-dimensional
model\cite{Stock_Woywod:1995} of pyrazine.

The remainder of this paper is organized as follows: In
Section~\ref{sec:theory}, after defining geometric properties of the exact
evolution operator, we discuss their breakdown in elementary methods and
recovery in the proposed symmetric compositions of the trapezoidal rule and
implicit midpoint methods. Next, we present the dynamic Fourier method for its
ease of implementation and the exponential convergence with the number of grid
points. Yet, the proposed integrators can be combined with
any other basis or grid representation. We conclude
Section~\ref{sec:results} by discussing the relationship between the molecular
Hamiltonians in the adiabatic and diabatic representations. In
Section~\ref{sec:results}, the convergence properties and conservation of
geometric invariants by various methods are analyzed numerically on a
two-surface NaI model\cite{Engel_Metiu:1989} in the adiabatic representation.
This system has a non-separable Hamiltonian due to an avoided
crossing between its potential energy surfaces and a corresponding region of
large nonadiabatic momentum coupling. Section~\ref{sec:conclusion}
concludes the paper.

\section{\label{sec:theory}Theory}

For a time-independent Hamiltonian $\hat{H}$, the time-dependent
Schr\"{o}dinger equation
\begin{equation}
i\hbar\frac{d\psi(t)}{dt}=\hat{H}\psi(t) \label{eq:tdse}%
\end{equation}
has the formal solution $\psi(t)=\hat{U}(t)\psi(0)$, where $\psi(0)$ is the
initial state and $\hat{U}(t)$ the so-called evolution operator. The exact
evolution operator
\begin{equation}
\hat{U}(t)=e^{-i\hat{H}t/\hbar} \label{eq:exact_evol_op}%
\end{equation}
is linear (in particular, independent of the initial state), reversible,
stable, and, moreover, conserves both the norm and energy of the quantum
state. Let us define and discuss these and other geometric properties of the
exact evolution operator because they are also desirable in approximate
numerical evolution operators $\hat{U}_{\text{appr}}(t)$.

\subsection{\label{subsec:exactprop}Geometric properties of the exact
evolution operator}

An operator $\hat{U}$ on a Hilbert space is said to \emph{preserve the norm}
$\Vert\psi\Vert:=\langle\psi|\psi\rangle^{1/2}$ if $\Vert\hat{U}\psi
\Vert=\Vert\psi\Vert$. For linear operators $\hat{U}$, preserving the norm is
equivalent to \emph{preserving the inner product},
\begin{equation}
\langle\hat{U}\psi|\hat{U}\phi\rangle\equiv\langle\psi|\hat{U}^{\dag}\hat
{U}\phi\rangle=\langle\psi|\phi\rangle, \label{eq:inner_prod_t_p_dt}%
\end{equation}
where $\hat{U}^{\dag}$ is the Hermitian adjoint of $\hat{U}$. The preservation
of inner product is, therefore, equivalent to the condition that $\hat
{U}^{\dagger}\hat{U}$ be the identity operator, or that
\begin{equation}
\hat{U}^{-1}=\hat{U}^{\dagger}. \label{eq:def_unitary_operator}%
\end{equation}
Such an operator $\hat{U}$ is said to be \emph{unitary}. The exact evolution
operator is unitary since $\hat{U}(t)^{\dag}=\exp(i\hat{H}%
t/\hbar)=\hat{U}(t)^{-1}$.

An operator $\hat{U}$ is said to be \emph{symplectic} if it preserves the
symplectic two-form $\omega(\psi,\phi)$, i.e., a nondegenerate skew-symmetric
bilinear form on the Hilbert space, if $\omega(\hat{U}\psi,\hat{U}\phi
)=\omega(\psi,\phi)$. In classical mechanics, conservation of the symplectic
two-form has many far-reaching consequences, one of which is Liouville's
theorem---the conservation of phase space volume. In quantum mechanics, a
symplectic two-form can be defined as\cite{book_Lubich:2008}
\begin{equation}
\omega(\psi,\phi):=-2\hbar\mathrm{Im}\langle\psi|\phi\rangle;
\label{eq:def_symplectic_two_form}%
\end{equation}
obviously, it is conserved if the inner product $\langle\psi|\phi\rangle$
itself is. The exact evolution operator is therefore symplectic.

The expectation value of \emph{energy} is conserved if the evolution operator
is unitary and commutes with the Hamiltonian:
\begin{align}
E(t)  &  =\langle\hat{H}\rangle_{\psi(t)}:=\langle\psi(t)|\hat{H}%
|\psi(t)\rangle\nonumber\\
&  =\langle\psi(0)|\hat{U}(t)^{\dagger}\hat{H}\hat{U}(t)|\psi(0)\rangle
\nonumber\\
&  =\langle\psi(0)|\hat{U}(t)^{\dagger}\hat{U}(t)\hat{H}|\psi(0)\rangle
\nonumber\\
&  =\langle\psi(0)|\hat{H}|\psi(0)\rangle=E(0). \label{eq:energy_conservation}%
\end{align}
The exact evolution operator is unitary, and because $\hat{U}%
(t)=\exp(-i\hat{H}t/\hbar)$ can be Taylor expanded into a convergent series
in powers of $\hat{H}$, $\hat{U}(t)$ also commutes with $\hat{H}$.
As a result, the exact evolution conserves energy.

An \emph{adjoint} $\hat{U}(t)^{\ast}$ of an evolution operator $\hat{U}(t)$ is
defined as its inverse taken with a reversed time step:
\begin{equation}
\hat{U}(t)^{\ast}:=\hat{U}(-t)^{-1}. \label{eq:def_adjoint}%
\end{equation}
An evolution operator is said to be \emph{symmetric} if it is equal to its own
adjoint:\cite{book_Hairer_Wanner:2006}
\begin{equation}
\hat{U}(t)^{\ast}=\hat{U}(t). \label{eq:def_symmetric}%
\end{equation}
An evolution is \emph{time-reversible} if a forward propagation for time $t$
is exactly cancelled by an immediately following backward propagation for the
same time, i.e., if\cite{book_Hairer_Wanner:2006}
\begin{equation}
\hat{U}(-t)\hat{U}(t)\psi(0)=\psi(0). \label{eq:def_reversible}%
\end{equation}
Time reversibility in quantum dynamics is, therefore, a direct consequence of
symmetry. The exact evolution operator is both symmetric and time-reversible
because $\hat{U}(t)^{\ast}=\exp(-i\hat{H}t/\hbar).$

Finally, the time evolution is said to be:

(i) \emph{stable}\cite{Auslander_Seibert:1964, book_Leimkuhler_Reich:2004,
book_Bhatia_George:2006} if for every $\epsilon>0,$ there exists
$\delta(\epsilon)>0$ such that
\begin{equation}
\Vert\psi(0)-\phi(0)\Vert<\delta\text{ implies }\Vert\psi(t)-\phi
(t)\Vert<\epsilon\text{ for all $t;$} \label{eq:stability_condition}%
\end{equation}

(ii) \emph{attracting}\cite{Auslander_Seibert:1964, book_Bhatia_George:2006}
if there exists a $\delta>0$ such that
\begin{equation}
\Vert\psi(0)-\phi(0)\Vert<\delta\text{ implies }\Vert\psi(t)-\phi
(t)\Vert\rightarrow0\text{ as }t\rightarrow\infty;
\label{eq:asymptotic_stability_condition}%
\end{equation}

(iii) \emph{asymptotically stable }if it is both stable and attracting.

\begin{figure}
[tbh]\includegraphics[scale=1]{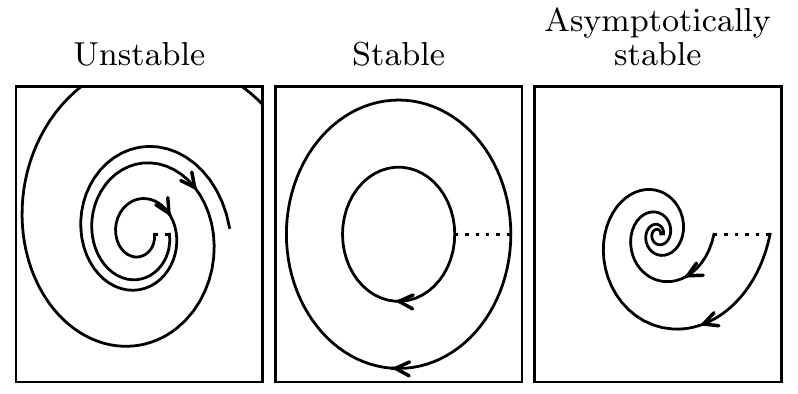}
\caption{Schematic representation of
stability conditions in Euclidean space; the distance between corresponding points
on the two curves (e.g., the tips of the arrows) is
analogous to a metric $\| \psi(t) - \phi(t) \|$ in the Hilbert space; the
dotted lines represent $\| \psi(0) - \phi(0) \|$.}\label{fig:picstability}
\end{figure}

These conditions are visualized in Fig.~\ref{fig:picstability}. The exact
evolution operator is stable but not asymptotically stable because, due to
norm conservation,
\begin{equation}
\Vert\psi(t)-\phi(t)\Vert=\Vert\psi(0)-\phi(0)\Vert.
\label{eq:exact_evol_op_stability}%
\end{equation}

\subsection{\label{subsec:lossprop}Loss of geometric properties by approximate
methods}

In approximate propagation methods, the state $\psi(t+\Delta t)$ at time
$t+\Delta t$, where $\Delta t$ is the numerical time step, is obtained from
the state $\psi(t)$ at time $t$ by applying an approximate time evolution
operator $\hat{U}_{\mathrm{appr}}(\Delta t)$. This operator is%
\begin{equation}
\hat{U}_{\mathrm{expl}}(\Delta t):=1-\frac{i}{\hbar}\Delta t\,\hat{H}
\label{eq:exp_euler}%
\end{equation}
in the \emph{explicit Euler} method and
\begin{equation}
\hat{U}_{\mathrm{impl}}(\Delta t):=\left(  1+\frac{i}{\hbar}\Delta t\,\hat
{H}\right)  ^{-1} \label{eq:imp_euler}%
\end{equation}
in the \emph{implicit Euler} method. Both Euler methods are of the first order
in the time step $\Delta t$, and both are neither unitary nor symplectic. Due
to their lack of unitarity, the methods do not conserve energy, even though
their evolution operators commute with the Hamiltonian. Neither method is
symmetric; in fact, they are adjoints of each other. Hence, neither method is
time-reversible. The explicit Euler method is unstable with the distance
between two wavefunctions diverging,
\begin{equation}
\Vert\psi_{\mathrm{expl}}(t)-\phi_{\mathrm{expl}}(t)\Vert\rightarrow
\infty\text{ as }t\rightarrow\infty, \label{eq:exp_eu_stability}%
\end{equation}
whereas the implicit Euler method is asymptotically stable.

The \emph{second-order differencing} method\cite{Kosloff_Kosloff:1983,
Leforestier_Kosloff:1991, Askar_Cakmak:1978} recovers symmetry by combining a
forward and backward steps of the explicit Euler method:%
\begin{equation}
\psi_{\mathrm{sod}}(t+\Delta t)-\psi_{\mathrm{sod}}(t-\Delta t)=-2\frac
{i}{\hbar}\Delta t\hat{H}\psi_{\mathrm{sod}}(t). \label{eq:sod_method}%
\end{equation}
The method can be also obtained directly from the time-dependent
Schr\"{o}dinger equation by using a finite-difference approximation
\begin{equation}
\frac{d\psi(t)}{dt}\approx\frac{\psi(t+\Delta t)-\psi(t-\Delta t)}{2\Delta t}.
\label{eq:second_order_t_derivative_wf}%
\end{equation}
While it is almost as simple as the explicit Euler method to implement, the
second order differencing has a higher order of accuracy and, in contrast to
the explicit Euler method, it is symmetric, time-reversible, and at least
\emph{conditionally stable}, meaning that it remains stable for sufficiently
small time steps $\Delta t$. The second order differencing does not conserve
the inner product, norm, energy, or symplectic two-form exactly. Yet, it
conserves quantities analogous to the inner product [see
Eq.~(\ref{eq:sod_inner_prod})], norm,\cite{Kosloff_Kosloff:1983,
Leforestier_Kosloff:1991} energy\cite{Kosloff_Kosloff:1983,
Leforestier_Kosloff:1991} [see Eq.~(\ref{eq:sod_energy_a})], and symplectic
two-form [see Eq.~(\ref{eq:sod_symplectic})]. The corresponding exact
quantities are conserved only up to the fourth order in $\Delta t$ (see Propositions 5 and 6 of Appendix~\ref{appendixa}).

The properties of the different methods are summarized in
Table~\ref{tab:properties}; a more thorough justification of these properties
is provided in Appendix~\ref{appendixa}. Although the explicit and implicit
Euler methods are not geometric, composing them in a specific way leads to
arbitrary-order integrators that preserve many important geometric properties
of the exact solution. Obviously, the compositions are applicable to systems
with non-separable Hamiltonians just like the elementary methods themselves.

\begin{table}
[pbh]%
\caption{Geometric properties and computational cost of various integrators. Cost is measured by the
number of Fourier transforms required per time
step (see Sec.~\ref{sec:mol_ham}).
$I$ is the number of iterations for the implicit step and $n=0,1,2,\ldots$ is the number of recursive
compositions. $C$ is the total number of composition steps per time step ($C = 3^{n}$  for
the triple jump\cite{Yoshida:1990, Suzuki:1990}, $C = 5^{n}$ for Suzuki's
fractals\cite{Suzuki:1990}).  $+$ denotes that the geometric
property of the exact evolution operator is preserved and $-$ denotes that it
is not. SOD stands for the second-order differencing and SO for  the
split-operator algorithm. }\label{tab:properties}\begin{ruledtabular}
\begin{tabular}{lccccccccc}
\multicolumn{1}{c}{}              & \multicolumn{1}{c}{Order} & \multicolumn{1}{c}{Unitary} & \multicolumn{1}{c}{Sympl-} &
\multicolumn{1}{c}{Commutes} & \multicolumn{1}{c}{Energy} & \multicolumn{1}{c}{Symm-} &
\multicolumn{1}{c}{Time-} & \multicolumn{1}{c}{Stable} & \multicolumn{1}{c}{Cost}\\
\multicolumn{1}{c}{}              & \multicolumn{1}{c}{} & \multicolumn{1}{c}{} & \multicolumn{1}{c}{ectic} &
\multicolumn{1}{c}{with $\hat{H}$} & \multicolumn{1}{c}{cons.} & \multicolumn{1}{c}{etric} &
\multicolumn{1}{c}{reversible} & \multicolumn{1}{c}{} & \\ \hline
\multicolumn{3}{l}{Elementary methods}                          &                & &         &               &            &  & \\ \hline
$1^{\mathrm{st}}$ order SO  & 1                         & $+$                             & $+$                              & $-$
& $-$                                  & $-$                            & $-$                                   & $+$ &
$2$
\\
Expl. Euler              & 1                         & $-$                             & $-$                              & $+$
& $-$                                  & $-$                            & $-$                                   & $-$ & $4D$
\\
Impl. Euler            & 1                         & $-$                             & $-$                              & $+$
& $-$                                  & $-$                            & $-$                                   & $+$ & $4D(2 + I)$
\\
SOD   & 2                         & $-$                             & $-$                             & $+$
& $-$                                  & $+$                            &  $+$                                   & $+$\footnotemark[1] &
$4D$
\\ \hline
\multicolumn{4}{l}{Recursively composable methods}                             & &         &               &            &  & \\ \hline
$2^{\mathrm{nd}}$ order SO & 2$(n+1)$                      & $+$                             & $+$                              & $-$
& $-$                                  & $+$                            & $+$                                   & $+$  &
$2C$
\\
Midpoint           & 2$(n+1)$                      & $+$                             & $+$                              & $+$
& $+$                                  & $+$                            & $+$                                   & $+$ & $4D(3 + I)C$
\\
Trapezoidal                & 2$(n+1)$                      & $+$                             & $+$                              & $+$
& $+$                                  & $+$                            & $+$                                   & $+$ & $4D(3 + I)C$
\\
\end{tabular}
\footnotetext{Stability holds for time steps that satisfy Eq.~(\ref{eq:sod_critical_timestep}).}
\end{ruledtabular}

\end{table}

\subsection{\label{subsec:compprop}Recovery of geometric properties by
composed methods}

Composing the explicit and implicit Euler methods, each for a time step
$\Delta t/2,$ yields a symmetric second-order method
(see Proposition
7 of Appendix~\ref{appendixa}). Depending on the order of composition, one
obtains either the \emph{trapezoidal rule}%
\begin{equation}
\hat{U}_{\mathrm{trap}}(\Delta t):=\hat{U}_{\mathrm{impl}}(\Delta t/2)\hat
{U}_{\mathrm{expl}}(\Delta t/2), \label{eq:trap_rule}%
\end{equation}
or \emph{implicit midpoint method}%
\begin{equation}
\hat{U}_{\mathrm{midp}}(\Delta t):=\hat{U}_{\mathrm{expl}}(\Delta t/2)\hat
{U}_{\mathrm{impl}}(\Delta t/2). \label{eq:imp_midpoint}%
\end{equation}

The trapezoidal rule is also known as the \emph{Crank-Nicolson method}%
,\cite{Crank_Nicolson:1947} although the latter frequently implies a
second-order finite-difference approximation to the spatial derivative in the
kinetic energy operator, whereas we use the dynamic Fourier method (see
Sec.~\ref{sec:dyn_fourier}), which has exponential convergence with grid
density (see Appendix~\ref{appendixb}).

Both the trapezoidal rule and implicit midpoint methods are Cayley
transforms\cite{book_Golub_Van_Loan:1996} of $(i\Delta t/2\hbar)\hat{H}$ and,
therefore, are unitary; in addition, both are second-order, symplectic,
symmetric, time-reversible, and stable regardless of the size of the time
step. Both methods also commute with the Hamiltonian, are energy conserving,
and can be further recursively composed to obtain arbitrary-order methods (see
Sec.~\ref{subsec:composition}). The summary of the properties is given in
Table~\ref{tab:properties} and a detailed justification provided in
Appendix~\ref{appendixa}.

It is necessary to stress that the geometric properties of the trapezoidal
rule and implicit midpoint method are only preserved if the implicit step,
which involves solving a set of linear equations, is executed exactly (or, in
practice, to machine accuracy). We solved the system of equations using the
generalized minimal residual method,\cite{book_Press_Flannery:2007,
Saad_Schultz:1986, book_Saad:2003} an iterative method based on Arnoldi
process.\cite{Arnoldi:1951, Saad:1980} It was an appropriate choice since the
matrix being inverted was not positive-definite, symmetric, skew-symmetric,
Hermitian, or skew-Hermitian, and therefore neither conjugate gradient nor
minimal residual method was applicable.\cite{book_Saad:2003} The initial guess
for the implicit step was approximated with the explicit Euler method since
for small time steps, the solutions from the explicit and implicit Euler
methods differ only by $(\Delta t/\hbar)^{2}\hat{H}^{2}|\psi(t)\rangle$.

\subsection{\label{subsec:composition}Symmetric composition schemes for
symmetric methods}

\begin{figure}
[pt]\includegraphics[scale=0.9,trim={0 0.5cm 0 0.8cm}]{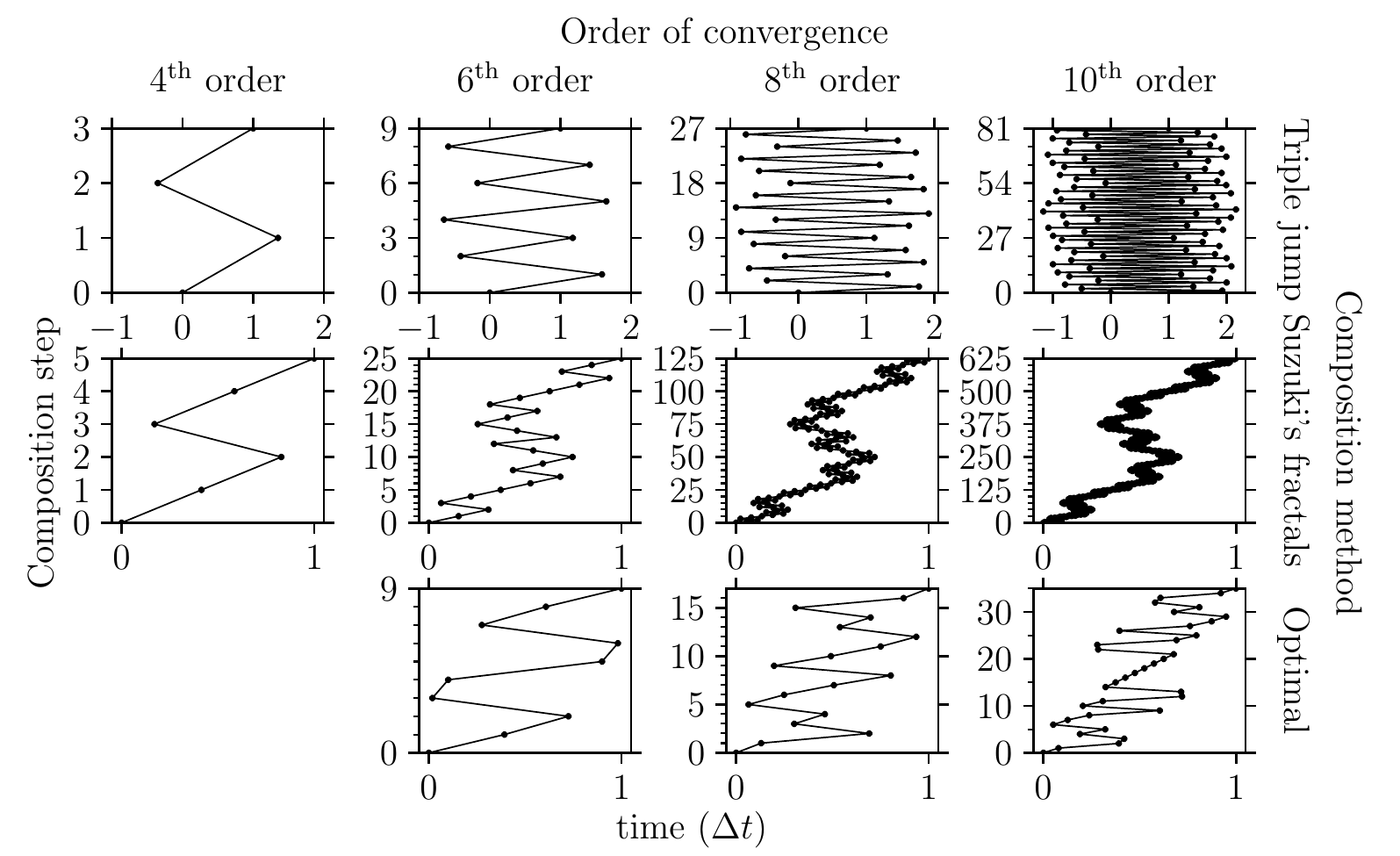}
\caption{Pictorial representation of recursive (triple
jump and Suzuki's fractals) and non-recursive ``optimal'' composition schemes. The
triple jump has $3^{\frac{p}{2} - 1}$ composition steps per time step, where
$p$ is the order of the method, whereas Suzuki's fractal has $5^{\frac{p}{2} - 1}$
composition steps.}\label{fig:triple_suzuki_optimal}
\end{figure}

Recursively composing symmetric methods with appropriately chosen time steps
leads to symmetric integrators of arbitrary
orders.\cite{book_Hairer_Wanner:2006,Suzuki:1990,Yoshida:1990} More precisely,
there exist a natural number $M$ and real numbers $\gamma_{n}$, $n=1,\ldots
,M$, called \emph{composition coefficients},
satisfying $\sum
_{n=1}^{M}\gamma_{n}=1$ and such that if $\hat{U}_{p}(\Delta t)$ is any
symmetric integrator of (necessarily even) order $p$, then%
\[
\hat{U}_{p+2}(\Delta t):=\hat{U}_{p}(\gamma_{M}\Delta t)\cdots\hat{U}%
_{p}(\gamma_{1}\Delta t)
\]
is a symmetric integrator of order $p+2$. The most common composition schemes
(see Fig.~\ref{fig:triple_suzuki_optimal}) are the triple
jump\cite{Cruetz_Gocksch:1989,Forest_Ruth:1990,Suzuki:1990,Yoshida:1990} with
$M=3$,%
\begin{equation}
\gamma_{1}=\frac{1}{2-2^{1/(p+1)}},\quad\gamma_{2}=-\frac{2^{1/(p+1)}%
}{2-2^{1/(p+1)}}, \label{eq:triple_jump}%
\end{equation}
and Suzuki's fractals\cite{Suzuki:1990} with $M=5$,
\begin{equation}
\gamma_{1}=\gamma_{2}=\frac{1}{4-4^{1/(p+1)}},\quad\gamma_{3}=-\frac
{4^{1/(p+1)}}{4-4^{1/(p+1)}}. \label{eq:suzukis_fractal}%
\end{equation}
The remaining coefficients are obtained from the relation $\gamma
_{M+1-n}=\gamma_{n}$, which expresses that both of these are \emph{symmetric
compositions}.

Because each triple jump is formed of three steps while each Suzuki's fractal
is composed of five steps, the $p^{\mathrm{th}}$-order integrator obtained
using Suzuki's fractals has $(5/3)^{\frac{p}{2}-1}$ times more composition
steps than the one obtained from the same symmetric second-order method using
the triple jump. Therefore, the $p^{\mathrm{th}}$-order method obtained from
Suzuki's fractals takes $(5/3)^{\frac{p}{2}-1}$ times longer to execute per
time step $\Delta t$ than does the method of the same order achieved through
the triple jump. Yet, the leading order error coefficient of the
$p^{\mathrm{th}}$-order integrator based on Suzuki's fractal is smaller
because the magnitude of each composition step is smaller in Suzuki's fractal.
Consequently, to achieve the same accuracy at a final time $t$, larger time
steps can be typically used for calculations using Suzuki's fractals compared
to those based on the triple jump.

Non-recursive composition schemes, which require fewer composition steps and
are also more efficient, have been obtained for various specific orders.
We will refer to these as \textquotedblleft optimal\textquotedblright%
\ schemes because they minimize the \textquotedblleft magnitude\textquotedblright\ of composition steps. The
magnitude of composition steps can be defined as either $\max_{n}%
|\gamma_{n}|$ or $\sum_{n=1}^{M}|\gamma_{n}|$. 
With either definition, Suzuki's fractal is the
optimal\ fourth-order scheme. The optimal\ sixth- and eighth-order
schemes,\cite{Kahan_Li:1997} found by Kahan and Li by minimizing $\max_{n}|\gamma
_{n}|$, have two more composition steps ($M=9$ and $M=17$, respectively) than the
minimum number possible for the respective order; the optimal\ tenth-order
scheme,\cite{Sofroniou_Spaletta:2005} obtained by Sofroniou and Spaletta by
minimizing $\sum_{n=1}^{M}|\gamma_{n}|$, has four more ($M=35$).

\textbf{Theorem}. All compositions of the trapezoidal rule or implicit
midpoint method are unitary, symplectic, stable, energy-conserving, and their
evolution operators commute with the Hamiltonian; all symmetric compositions
are symmetric and therefore time-reversible.

\textbf{Proof. }We prove the theorem in much greater generality. Indeed, a
composition of \emph{any} unitary operators $\hat{U}_{1}$ and $\hat{U}_{2}$ is
unitary since%
\[
(\hat{U}_{2}\hat{U}_{1})^{\dag}=\hat{U}_{1}^{\dag}\hat{U}_{2}^{\dag}=\hat
{U}_{1}^{-1}\hat{U}_{2}^{-1}=(\hat{U}_{2}\hat{U}_{1})^{-1}.
\]
A composition of any symplectic operators is symplectic since%
\[
\omega(\hat{U}_{2}\hat{U}_{1}\psi,\hat{U}_{2}\hat{U}_{1}\phi)=\omega(\hat
{U}_{1}\psi,\hat{U}_{1}\phi)=\omega(\psi,\phi).
\]
Proposition 3 of Appendix~\ref{appendixa} shows that a composition of any
operators commuting with the Hamiltonian commutes with the Hamiltonian. A
composition of any energy-conserving operators conserves energy since%
\[
\langle\hat{H}\rangle_{\hat{U}_{2}\hat{U}_{1}\psi}=\langle\hat{H}\rangle
_{\hat{U}_{1}\psi}=\langle\hat{H}\rangle_{\psi}.
\]
However, a composition of two symmetric operators is, in general, not
symmetric:%
\[
(\hat{U}_{2}\hat{U}_{1})^{\ast}=\hat{U}_{1}^{\ast}\hat{U}_{2}^{\ast}=\hat
{U}_{1}\hat{U}_{2}\neq\hat{U}_{2}\hat{U}_{1}.
\]
It is symmetric if the two operators commute or if it is a symmetric
composition, e.g.,%
\[
(\hat{U}_{1}\hat{U}_{2}\hat{U}_{1})^{\ast}=\hat{U}_{1}^{\ast}\hat{U}_{2}%
^{\ast}\hat{U}_{1}^{\ast}=\hat{U}_{1}\hat{U}_{2}\hat{U}_{1}.
\]
Finally, a composition of time-reversible operators is not necessarily
time-reversible since%
\[
\hat{U}_{2}(-\Delta t_{2})\hat{U}_{1}(-\Delta t_{1})\hat{U}_{2}(\Delta
t_{2})\hat{U}_{1}(\Delta t_{1})=\hat{U}_{2}(\Delta t_{2})^{-1}\hat{U}%
_{1}(\Delta t_{1})^{-1}\hat{U}_{2}(\Delta t_{2})\hat{U}_{1}(\Delta t_{1}%
)\neq1.
\]
The composition is time-reversible if the two operators commute or if it is a
symmetric composition, e.g.,%
\begin{align*}
&  \hat{U}_{1}(-\Delta t_{1})\hat{U}_{2}(-\Delta t_{2})\hat{U}_{1}(-\Delta
t_{1})\hat{U}_{1}(\Delta t_{1})\hat{U}_{2}(\Delta t_{2})\hat{U}_{1}(\Delta
t_{1})\\
&  =\hat{U}_{1}(\Delta t_{1})^{-1}\hat{U}_{2}(\Delta t_{2})^{-1}\hat{U}%
_{1}(\Delta t_{1})^{-1}\hat{U}_{1}(\Delta t_{1})\hat{U}_{2}(\Delta t_{2}%
)\hat{U}_{1}(\Delta t_{1})=1.
\end{align*}

\subsection{\label{sec:dyn_fourier}Dynamic Fourier method}

To propagate the wavepacket using the explicit or implicit Euler method, or
one of their compositions (see Sec.~\ref{subsec:lossprop}%
--\ref{subsec:composition}), only the action of the Hamiltonian operator
$\hat{H}$ on $\psi(t)$ is required provided that the implicit steps are solved
iteratively. The dynamic Fourier method\cite{Feit_Steiger:1982,
Kosloff_Kosloff:1983a, Kosloff_Kosloff:1983, book_Tannor:2007} is an efficient
approach to compute $f(\hat{x})\psi(t)$, where $f(\hat{x})$ is an arbitrary
function of $\hat{x}$, which denotes either the nuclear position ($\hat{q}$)
or momentum ($\hat{p}$) operator. Each action of $f(\hat{x})$ on $\psi(t)$ is
evaluated in the $x$-representation (in which $\hat{x}$ is diagonal) by a
simple multiplication, after Fourier-transforming $\psi(t)$ to change the
representation if needed. On a grid of $N$ points, $f(\hat{x})\psi(t)$ is
evaluated as $f(x_{i})\psi(x_{i},t)$, $1\leq i\leq N$, where $\psi(x,t)$ is
the wavepacket in the $x$-representation and $x_{i}$ are either the position
or momentum grid points.

\subsection{\label{sec:mol_ham}Molecular Hamiltonian in the adiabatic basis}

The molecular Hamiltonian in the adiabatic basis can be expressed as
\begin{equation}
\hat{\mathbf{H}}=\frac{1}{2}[\hat{p}\bm{1}-i\hbar\mathbf{F}(\hat{q})]^{\dag
}\cdot m^{-1}\cdot\lbrack\hat{p}\bm{1}-i\hbar\mathbf{F}(\hat{q})]+\mathbf{V}(\hat{q}),\label{eq:adiab_mol_ham}\end{equation}
where $m$ is the diagonal $D\times D$ nuclear mass matrix, $D$ the number of
nuclear degrees of freedom, $\mathbf{V}$ the diagonal $S\times S$
potential energy matrix, $S$ the number of considered
electronic states, and $\mathbf{F}$ the nonadiabatic coupling vector
(more precisely, a $D$-vector of $S\times S$
matrices). In Eq.~(\ref{eq:adiab_mol_ham}), the dot $\cdot$ denotes the
matrix product in nuclear $D$-dimensional vector space, the hat $\hat{}$
represents a nuclear operator, and the \textbf{bold} font indicates an
electronic operator, i.e., an $S\times S$ matrix. Using the dynamic Fourier
method, each evaluation of the action of $\hat{\mathbf{H}}$ on a molecular
wavepacket $\bm{\psi}(t)$, which now becomes an $S$-component vector of
nuclear wavepackets (one on each surface), involves $4D$ changes of the
wavepacket's representation.

In two-state models (i.e., for $S=2$), it is possible to obtain
$\hat{\mathbf{H}}$ in the adiabatic representation analytically from the one
in the diabatic representation,\cite{Sadygov_Yarkony:1998, Baer_Englman:1992,
Hobey_McLachlan:1960}
\begin{equation}
\hat{\mathbf{H}}_{\text{diab}}=\frac{1}{2}\hat{p}^{T}\cdot m^{-1}\cdot\hat
{p}\,\mathbf{1}+\mathbf{W}(\hat{q}),
\end{equation}
in which $\mathbf{W}(q)$ is the (real) diabatic potential energy matrix and in
which the nonadiabatic vector couplings vanish. The adiabatic potential
energy matrix $\mathbf{V}(q)$ is obtained by diagonalizing its diabatic analog
$\mathbf{W}(q)$,
\begin{equation}
\mathbf{V}(q)=\mathbf{O}(q)^{T}\mathbf{W}(q)\mathbf{O}%
(q),\label{eq:pot_diab2adiab}%
\end{equation}
and the molecular wavepacket in the adiabatic basis $\bm{\psi}(t)$ is
obtained from its counterpart in the diabatic basis $\bm{\psi}_{\mathrm{diab}%
}(t)$ as
\begin{equation}
\bm{\psi}(t)=\mathbf{O}(\hat{q})^{T}\bm{\psi}_{\mathrm{diab}}(t),
\label{eq:psi_diab2adiab}%
\end{equation}
using an orthogonal matrix
\begin{equation}
\mathbf{O}(q)=\frac{1}{\sqrt{W_{12}(q)^{2}+\Delta(q)^{2}}}%
\begin{pmatrix}
W_{12}(q), & -\Delta(q)\\
\Delta(q), & W_{12}(q)
\end{pmatrix}
,\label{eq:pot_diab2adiab_orthogonal_transformation}%
\end{equation}
with $\Delta(q)=V_{1}(q)-W_{11}(q)=-[V_{2}(q)-W_{22}(q)]$. The two adiabatic
energies are given by%
\[
V_{1,2}(q)=\bar{W}(q)\pm\sqrt{\lbrack\Delta W(q)/2]^{2}+W_{12}(q)^{2}},
\]
where $\Delta W:=W_{22}-W_{11}$ and $\bar{W}:=(W_{11}+W_{22})/2$. Finally, the
transformed momentum operator is
\begin{equation}
\mathbf{O}(\hat{q})^{T}\hat{p}\,\mathbf{O}(\hat{q})=\hat{p}\bm{1}-i\hbar
\mathbf{O}(\hat{q})^{T}\mathbf{O}^{\prime}(\hat{q}%
).\label{eq:diab2adiab_transformed_p}%
\end{equation}
By comparing with Eq.~(\ref{eq:adiab_mol_ham}), we see that, in the adiabatic
basis, the nonadiabatic coupling vector satisfies
$\mathbf{F}(\hat{q})=\mathbf{O}(\hat{q})^{T}\mathbf{O}^{\prime}(\hat{q}%
)$; in particular,
\begin{align}
F_{11}(q)  &  =F_{22}(q)=0,\nonumber\\
F_{12}(q)  &  =-F_{21}(q)=\frac{W_{12}^{\prime}(q)\Delta(q)-W_{12}%
(q)\Delta^{\prime}(q)}{W_{12}(q)^{2}+\Delta(q)^{2}}.\label{eq:nonadiab_c}%
\end{align}

\section{\label{sec:results}Numerical examples}

To test the geometric and convergence properties of the integrators presented
in Sections~\ref{subsec:lossprop}--\ref{subsec:composition}, we used these
integrators to simulate the nonadiabatic quantum dynamics in a two-surface
model\cite{Engel_Metiu:1989} of the NaI molecule.
This one-dimensional
model, motivated by the experiment by Mokhtari \textit{et al.},\cite{Mokhtari_Zewail:1990} has two electronic states,
and therefore an analytical transformation between diabatic and adiabatic
representations is available (see Sec.~\ref{sec:mol_ham}). This allowed us to compare the proposed
integrators, applied in the adiabatic basis, with the split-operator
algorithm, which can only be used in the diabatic basis. Such a rigorous
comparison would only be possible
for a two-surface model potential because the split-operator algorithm
requires the diabatization of the Hamiltonian formulated in the adiabatic
representation and this cannot be done exactly for higher-dimensional
\textit{ab initio} potential energy surfaces
with more electronic
states.

\begin{figure}
[tbh]\includegraphics[scale=1.0]{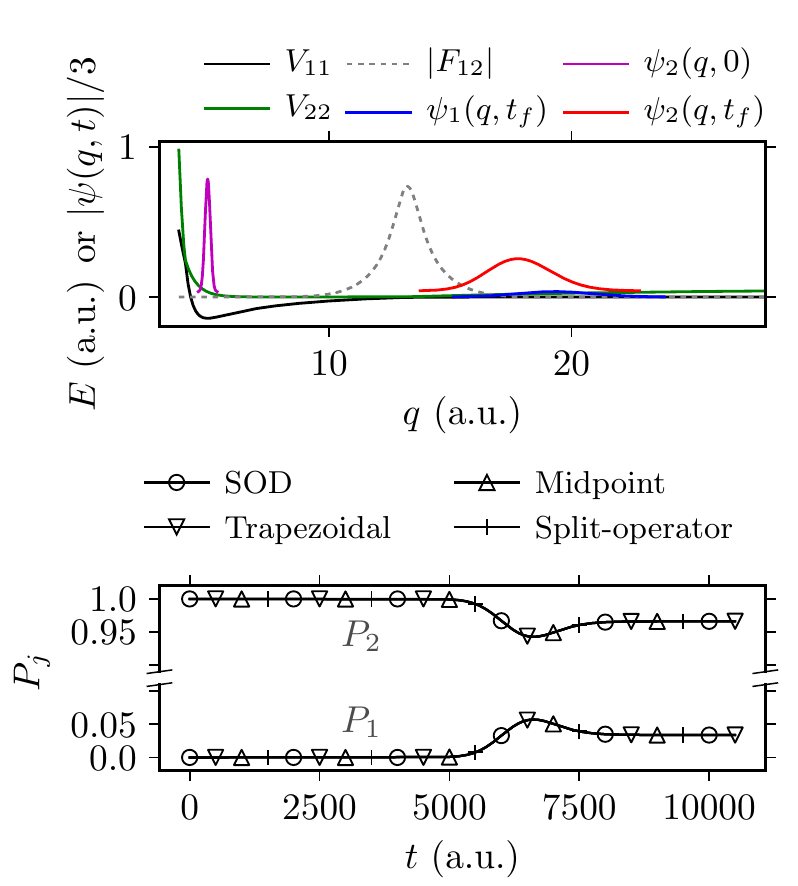}
\caption{\label{fig:P_and_psi}Nonadiabatic dynamics of NaI. Top: Adiabatic potential
energy surfaces with the initial and final nuclear wavepacket components in the ground and excited
adiabatic electronic states [Because the initial molecular wavepacket was in the excited state,
its component $\psi_{1}(q,t=0) \equiv 0$ is not shown.].
Bottom: Ground- and excited-state populations of NaI computed with four
different second-order methods: SOD stands for the second-order differencing. The populations were propagated
with a time step $\Delta t = 0.01$ a.u., i.e., much more frequently than the markers suggest. The
small time step guaranteed wavepacket convergence errors below $\approx 10^{-6}$ in all methods.}
\end{figure}

Before the electronic excitation, the NaI molecule was assumed to be in the
ground vibrational eigenstate of a harmonic fit to the ground-state potential
energy surface at the equilibrium geometry. This vibrational wavepacket was
then lifted to the excited-state surface, in order to obtain an initial
Gaussian wavepacket ($q_{0}=4.9889$ a.u., $p_{0}=0$ a.u., $\sigma
_{0}=0.110436$ a.u.) for the nonadiabatic dynamics. This use of the sudden
approximation assumes an impulsive excitation,
i.e., the simultaneous validity of the time-dependent perturbation theory and Condon and 
ultrashort pulse approximations during the excitation
process. After that, the nonadiabatic dynamics was performed by solving the
time-dependent Schr\"{o}dinger equation
using the dynamic Fourier
method (see Sec.~\ref{sec:dyn_fourier}) on a uniform grid with $2048$ points
between $q=3.8$ a.u. and $q=47.0$ a.u.; Appendix~\ref{appendixb} shows
wavepacket represented on such a grid is converged for the duration of the
dynamics. A long-enough propagation time, $t_{f}=10500$ a.u., was chosen so
that the wavepacket traverses the avoided crossing and simultaneously
witnesses the change of the nature of the excited adiabatic state from
covalent to ionic. The top panel of Fig.~\ref{fig:P_and_psi} shows the two
adiabatic potential energy surfaces as well as the initial wavepacket at $t=0$
and the final wavepacket at $t=t_{f}$. The population dynamics of NaI,
displayed in the bottom panel of Fig.~\ref{fig:P_and_psi}, shows that after
crossing the region of highest nonadiabatic coupling, most of the wavepacket
remains in the bound excited adiabatic state, while a small population
transfer occurs to the dissociative ground state. The figure also confirms
that the converged populations obtained with different integrators agree on
the scale visible in the figure; in particular, the results obtained with
integrators designed for the adiabatic basis agree with each other and also
with the result of the split-operator algorithm in the diabatic basis.

To compare various integrators quantitatively, it is essential to
\textquotedblleft zoom in\textquotedblright\ and inspect the convergence error
at the final time $t_{f}$; after all, the dynamic Fourier
method\cite{Feit_Steiger:1982, Kosloff_Kosloff:1983, book_Tannor:2007} is
expected to describe the wavepacket with a high degree of accuracy. In our
setting, the convergence error at time $t$ is defined as the $L_{2}$-norm
error $\left\Vert \psi_{\Delta t}(t)-\psi_{\Delta t/2}(t)\right\Vert $, where
$\psi_{\Delta t}(t)$ denotes the wavepacket propagated with a time step
$\Delta t$. We omit the split-operator algorithm, which served as a benchmark
in Fig.~\ref{fig:P_and_psi}, from the following analysis because this
algorithm is not applicable to time propagation in the adiabatic
representation. Note, however, that for separable Hamiltonians, such as the
nonadiabatic Hamiltonian in the diabatic basis, the split-operator algorithms
are more efficient than the present integrators of the corresponding order
(see Table I and Paper II).

\begin{figure}
[pbh]\includegraphics[scale=0.8]{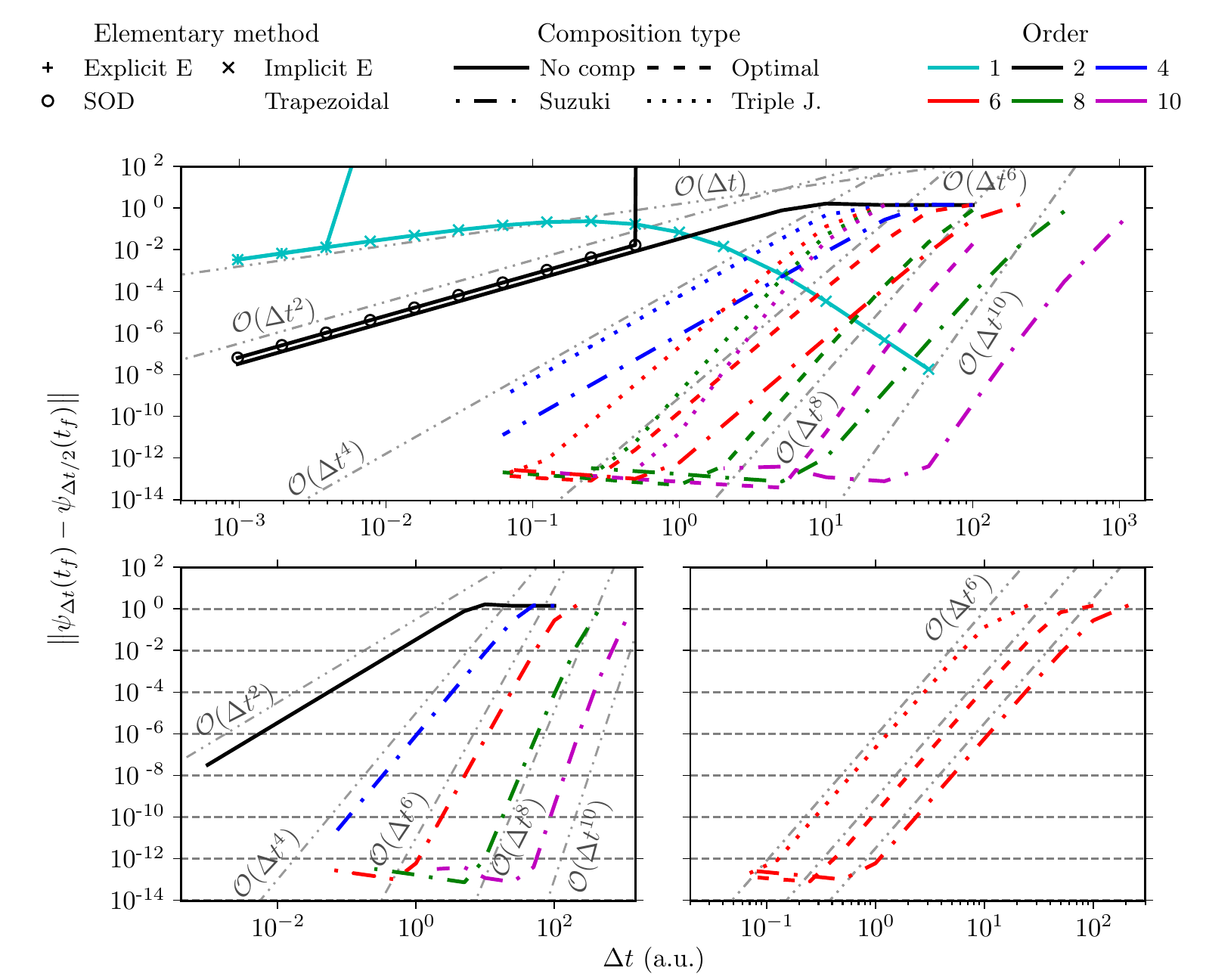}
\caption{Convergence of the molecular wavefunction as a function of the time step. Gray lines were added to guide
the eye. Top: all discussed methods; bottom left: methods composed through
Suzuki's fractals, bottom right: sixth-order methods.
}\label{fig:convergence}
\end{figure}

Figure~\ref{fig:convergence} plots the convergence error as a function of the
time step and confirms, for each algorithm, the asymptotic order of
convergence predicted in Sections~\ref{subsec:lossprop}%
--\ref{subsec:composition}. Recall that the trapezoidal rule and implicit
midpoint method are obtained by composing the explicit and implicit Euler
methods, and that the higher order methods are obtained from the trapezoidal
rule or implicit midpoint method using the triple jump, Suzuki's fractal, or
optimal composition. The top panel of Fig.~\ref{fig:convergence} compares the
convergence of all methods, while, for clarity, the bottom left-hand panel
only compares the different orders of composition for the Suzuki's fractal and
the bottom right-hand panel compares different composition schemes with the
sixth order of convergence. (In Fig.~\ref{fig:convergence} and all following
figures, we have omitted the results of the implicit midpoint method and of
its compositions because they overlap almost perfectly with the corresponding
results for the trapezoidal rule.) It is clear that, for a given order of
convergence, the prefactor of the error is the largest for the triple jump
composition, \cite{Yoshida:1990, Suzuki:1990} intermediate for the optimally
composed\cite{Kahan_Li:1997, Sofroniou_Spaletta:2005} method, and smallest for
Suzuki's fractal\cite{Suzuki:1990} composition. To guarantee the correct order
of convergence of all composed methods, the composed elementary second-order
method must be exactly symmetric, which requires that the systems of linear
equations arising from implicit steps must be solved numerically exactly.

In Section~\ref{subsec:lossprop}, we mentioned the instability of the explicit
Euler method and the conditional stability of the second-order differencing.
Both properties are reflected in the top panel of Fig.~\ref{fig:convergence}
in the divergence of the errors of the two methods for large time steps. The
critical time step for the second-order differencing is $\Delta t\approx0.5$
a.u., whereas the explicit Euler method is unstable regardless of $\Delta t$
but the effect of instability is more visible for larger time steps. In
contrast, the trapezoidal rule, implicit midpoint method, and their
compositions are stable, but implicit, and, therefore, require the solution of
systems of linear equations. These methods could not be used beyond a certain
time step ($\max_{n}| \gamma_{n} |\Delta t\approx100$ a.u. for both the
trapezoidal rule and implicit midpoint method) because the iterative
generalized minimal residual algorithm did not converge for very large $\Delta
t$.

\begin{figure}
[pbh]\includegraphics[scale=0.8]{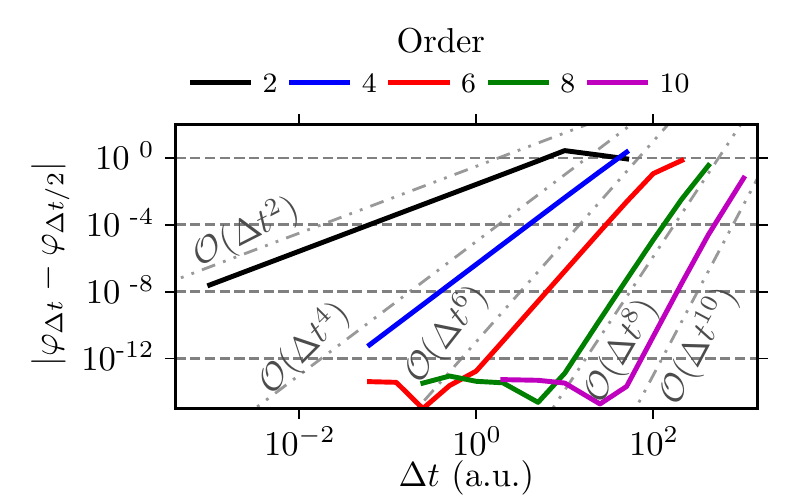} \caption{Convergence of the phase as a function of the time step. All higher-order integrators were obtained through Suzuki's fractal composition. Gray lines were added to guide the eye.\label{fig:convergence_phase}}
\end{figure}

Convergence of the wavepacket's phase, which is very important,
e.g., in the evaluation of spectra, is shown in
Fig.~\ref{fig:convergence_phase}. As a measure of the convergence error of the
phase, we use $\lvert\varphi_{\Delta t}-\varphi_{\Delta t/2} \rvert$, where $\varphi_{\Delta t}:=\mathrm{arg}[\psi_{\Delta t}(q_{\mathrm{\max}},t_{f})]$ and $q_{\mathrm{\max}}:=\mathrm{arg\,max}_{q}(\lvert\psi_{\Delta
t}(q,t_{f})\rvert)$, i.e., $\varphi_{\Delta t}$ is the phase of the
wavefunction propagated with time step $\Delta t$ at the position
$q_{\mathrm{max}}$, for which the amplitude of the wavefunction achieves its
maximum. Note that the order of convergence is identical to that of the
wavepacket.

\begin{figure}
[pbh]\includegraphics[scale=0.8 ]{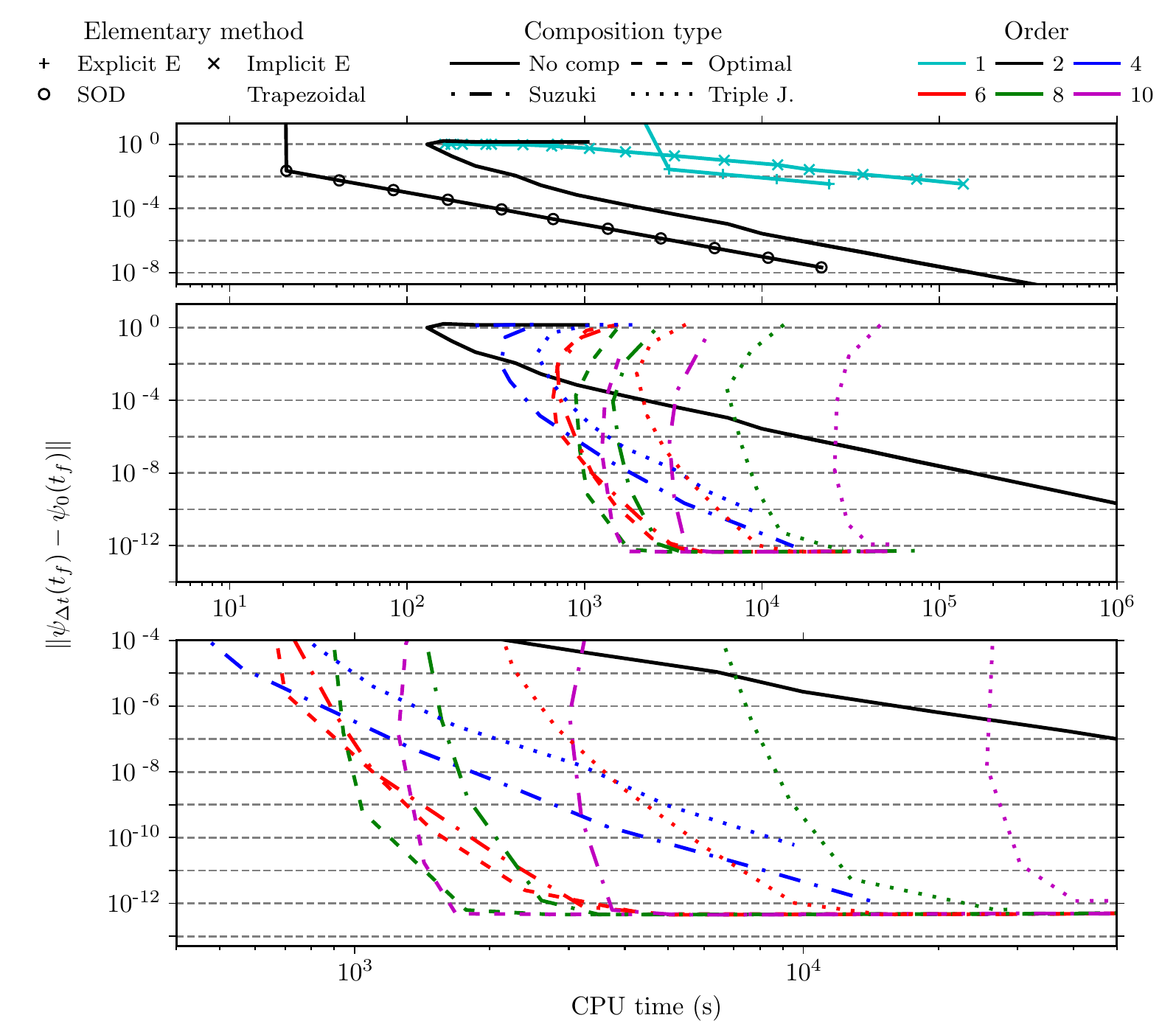} \caption{Efficiency of various
methods shown using the dependence of the convergence error on the computational (CPU) time.
Results of the elementary trapezoidal rule were extrapolated using the
line of best fit to highlight the speedup achieved with higher-order compositions.
Top: elementary methods; middle: trapezoidal rule and its compositions;
bottom: detail of the middle panel.}\label{fig:efficiency}
\end{figure}

The efficiency of an algorithm cannot be judged solely from the convergence
error for a given time step $\Delta t$ because the number of composition steps
depends on the composition scheme, and, indeed, grows exponentially for the
triple-jump and Suzuki's fractal compositions. Figure~\ref{fig:efficiency}, 
therefore, displays the convergence error $\left\Vert \psi_{\Delta t}%
(t)-\psi_{0}(t)\right\Vert $, where $\psi_{0}(t)$ is the wavepacket propagated
using the optimally composed $10^{\mathrm{th}}$-order trapezoidal
rule with an infinitesimal time step (in practice,
$\Delta t=0.01$
a.u.), as a function of the computational (CPU)\ time. Among the elementary
first- and second-order methods, compared in the top panel of
Fig.~\ref{fig:efficiency}, the second-order differencing, which does not
require the solution of a system of linear equations, is the most efficient.
Comparison of the geometric integrators based on the trapezoidal rule in the
middle and bottom panels of Fig.~\ref{fig:efficiency} shows that the
fourth-order Suzuki composition takes less CPU time to achieve convergence
error as high as $10^{-2}$ than does the elementary trapezoidal rule (i.e.,
Crank-Nicolson method). To reach errors below $10^{-2}$, it is already more
efficient to use the higher-order integrators. For a more dramatic example,
note that the CPU time required to reach an error of $10^{-10}$ is roughly
$1000$ times longer for the original trapezoidal rule than for its optimal
$8^{\mathrm{th}}$-order composition. (In Paper II, we confirm that
this gain in efficiency holds in higher dimensions by applying the
compositions of the trapezoidal rule to the nonadiabatic dynamics in a
three-dimensional model of pyrazine in the diabatic
representation.\cite{Stock_Woywod:1995}) The bottom panel of
Fig.~\ref{fig:efficiency} confirms the prediction that the optimal
compositions are the most efficient among composition methods of the same
order. Finally, note that the dependence of CPU\ time on the error in
Fig.~\ref{fig:efficiency} is not monotonous for the integrators with implicit
steps because the convergence of the numerical solution to the system of
linear equations required more iterations for larger time steps; as a result,
both the error and CPU time increased for time steps larger than a certain
critical value (see Fig.~\ref{fig:efficiency}).

\begin{figure}
[ptb]\includegraphics[scale=0.9]{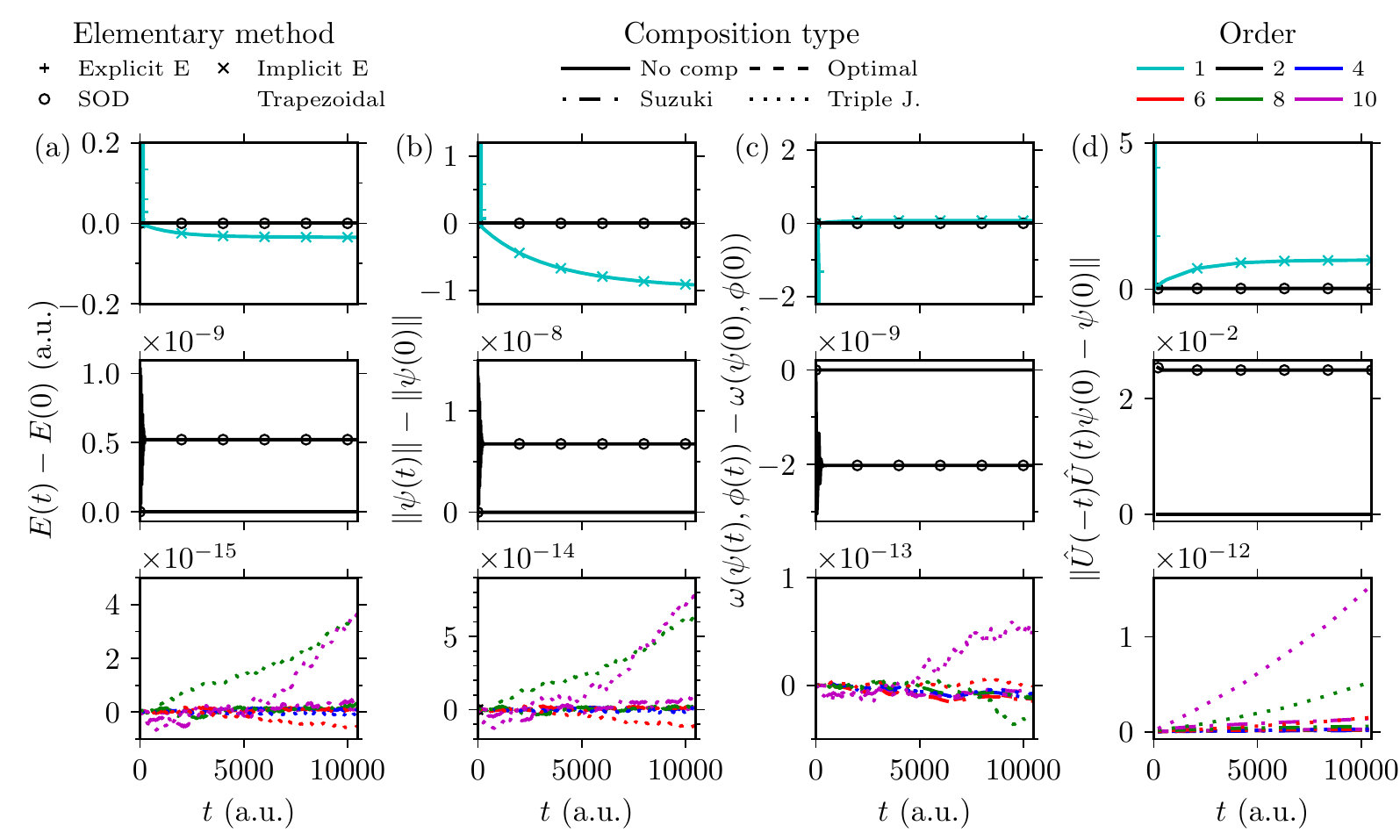}
\caption{Conservation of (a) energy, (b) norm, (c) symplectic two-form, and (d) time reversibility by the Euler and second-order differencing methods (top panels), elementary second-order methods (middle panels), and composed methods (bottom panels).
$\phi(0)$ is a complex Gaussian wavepacket with $q_{0} = 5.05$ a.u., $p_{0} = 2.5$ a.u., and $\sigma_{0}$ identical
to that of $\psi(0)$. As a consequence, $\omega(\psi(0), \phi(0))$ is nonzero.
For the Euler and second-order differencing
methods the time step was $\Delta t = 0.5$ a.u. to ensure the stability of the second-order differencing.
For all other methods, ten times larger time step ($\Delta t = 5$ a.u.) was used to highlight that the exact conservation of invariants is independent of the accuracy of the wavefunction itself.\label{fig:geom_prop}}
\end{figure}

Besides increased efficiency, another benefit of the algorithms based on the
composition of the trapezoidal rule or implicit midpoint method is the
conservation of the geometric properties of the exact evolution operator.
Conservation of the energy, norm, symplectic two-form, and time
reversibility by the trapezoidal rule and their compositions is demonstrated
in Fig.~\ref{fig:geom_prop}. Time reversibility is measured by the distance of
an initial state $\psi(0)$ from $\hat{U}(-t)\hat{U}(t)\psi(0)$, i.e., a state
propagated first forward in time for time $t$ and then backward in time for
time $t$. The tiny residual errors ($<2\cdot10^{-12}$ in all cases) of the
invariants result from accumulated numerical errors of the fast Fourier
transform and generalized minimal residual algorithm. In contrast, the
second-order differencing conserves energy, norm, and symplectic two-form only
approximately with much larger, $O(\Delta t^{4})$ errors (see Propositions 5
and 6 of Appendix~\ref{appendixa}). Although the second-order differencing is
time-reversible in theory, its practical implementation is not. [For the
second-order differencing to be exactly time-reversible, the wavepackets at
time $t=0$ and $t=-\Delta t$ would have to be known exactly before the start
of the simulation. However, because only $\psi(0)$ is typically available,
$\psi(-\Delta t)$ must be approximated with explicit methods such as the
second-order Runge--Kutta scheme.\cite{Kosloff_Kosloff:1983}] None of the four
geometric properties or analogous quantities is conserved by the Euler
methods. The explicit Euler method is unstable regardless of $\Delta t$, and
will, for long enough times $t_{f}$, result in a norm divergent to infinity [see Fig.~\ref{fig:geom_prop}(b), top panel] even for very
small $\Delta t$, implying that also the wavefunction will have an error
increasing beyond any bound. As for the implicit Euler method, its error of
the norm converges to $-1$ because $\Vert\psi_{\mathrm{impl}}(t)\Vert
\rightarrow0$ as $t\rightarrow\infty$ [see Fig.~\ref{fig:geom_prop}(b), top panel].

\section{\label{sec:conclusion}Conclusion}

We have described geometric integrators for nonadiabatic quantum dynamics in
the adiabatic representation, in which the popular split-operator algorithms
cannot be used due to nonseparability of the Hamiltonian into a kinetic and
potential terms. The proposed methods are based on the symmetric composition
of the trapezoidal rule or implicit midpoint method, and as a result, are
symmetric, stable, conserve the energy exactly and, in addition, are exactly
unitary, symplectic, and time-reversible. We have shown that unlike the
original trapezoidal rule or implicit midpoint method, which are only
second-order, their recursive symmetric compositions can achieve accuracy of
arbitrary even order in the time step.

We have proven all these properties analytically as well as demonstrated them
numerically on a two-surface model of NaI photodissociation. As expected, the
higher-order integrators significantly sped up calculations when higher
accuracy was required. For example, even to achieve a moderate wavefunction
convergence error of $10^{-5}$, tenfold reduction in the computational time
was observed by using higher-order methods compared to the elementary
trapezoidal rule. It is probable that Chebyshev\cite{Tal-Ezer_Kosloff:1984,
Tal-Ezer:1989} and short iterative Lanczos schemes\cite{Lanczos:1950,
Park_Light:1986} would be more efficient in this and other typical systems,
but these methods do not conserve exactly all the invariants conserved by the
described geometric integrators.

Finally, we hope that the ability to run \textquotedblleft
geometric\textquotedblright\ quantum molecular dynamics in the adiabatic
representation will be useful especially in conjunction with potential energy
surfaces obtained from \emph{ab initio} electronic structure calculations
because this will avoid the tedious diabatization process necessary for the
applicability of the split-operator algorithm.

The authors acknowledge the financial support from the European Research
Council (ERC) under the European Union's Horizon 2020 research and innovation
programme (grant agreement No. 683069 -- MOLEQULE), and thank Nikolay Golubev
and Rob Parrish for useful discussions.

\appendix

\section{\label{appendixa}Geometric properties of various integrators}

To shorten formulas, we set $\hbar=1$ and denote the increment $\Delta t$ with
$\epsilon$ throughout the Appendix. The $\hbar$ can be reintroduced by
replacing each occurrence of $t$ with $t/\hbar$ (and $\epsilon$ with
$\epsilon/\hbar$). To analyze geometric properties of various integrators, we
will use the following operator identities:

\textbf{Proposition 1. }Let $\hat{A}$ and $\hat{B}$ be invertible operators on
a Hilbert space, and let $\hat{A}^{\dag}$ and $\hat{B}^{\dag}$ their Hermitian
adjoints. Then both $\hat{A}^{\dag}$ and $\hat{A}\hat{B}$ are invertible, and
the following identities hold:%
\begin{align}
(\hat{A}^{\dag})^{-1}  &  =(\hat{A}^{-1})^{\dag},\label{eq:conj_inv}\\
(\hat{A}\hat{B})^{-1}  &  =\hat{B}^{-1}\hat{A}^{-1},\label{eq:prod_inv}\\
(\hat{A}\hat{B})^{\dag}  &  =\hat{B}^{\dag}\hat{A}^{\dag},
\label{eq:prod_conj}\\
(\hat{A}^{\dag})^{\dag}  &  =(\hat{A}^{-1})^{-1}=\hat{A}.
\end{align}

The first property expresses the compatibility of the inverse and Hermitian
adjoint operations, while the last three properties express that these two
operations are involutive antiautomorphisms on the group of invertible
operators. All four properties are easy to prove in finite-dimensional
spaces;\cite{book_Halmos:1942} the proofs for infinite-dimensional spaces can
be found in textbooks on advanced linear algebra or functional
analysis.\cite{book_Halmos:1951}

\textbf{Proposition 2. }Let $\hat{A}$ and $\hat{B}$ be commuting operators on
a vector space, i.e., $[\hat{A},\hat{B}]:=\hat{A}\hat{B}-\hat{B}\hat{A}=0$. If
$\hat{A}$ is invertible, then $[\hat{A}^{-1},\hat{B}]=0$. If both $\hat{A}$
and $\hat{B}$ are invertible, then $[\hat{A}^{-1},\hat{B}^{-1}]=0$.

The first statement follows from the sequence of identities%
\[
\hat{A}^{-1}\hat{B}=\hat{A}^{-1}\hat{B}\hat{A}\hat{A}^{-1}=\hat{A}^{-1}\hat
{A}\hat{B}\hat{A}^{-1}=\hat{B}\hat{A}^{-1}.
\]
The second statement follows from the first by applying it twice, the second
time for $\hat{A}:=\hat{B}$ and $\hat{B}:=\hat{A}^{-1}$, or directly from%
\[
\hat{A}^{-1}\hat{B}^{-1}=(\hat{B}\hat{A})^{-1}=(\hat{A}\hat{B})^{-1}=\hat
{B}^{-1}\hat{A}^{-1}%
\]
by using property (\ref{eq:prod_inv}).

\textbf{Proposition 3}. Let $\hat{A}$, $\hat{B}$, and $\hat{H}$ be operators
on a vector space. If $[\hat{H},\hat{A}]=[\hat{H},\hat{B}]=0$, then $[\hat
{H},\hat{A}\hat{B}]=0$.

This follows immediately from the identity $[\hat{H},\hat{A}\hat{B}]=\hat
{A}[\hat{H},\hat{B}]+[\hat{H},\hat{A}]\hat{B}$.

\subsection{\label{localerror}Local error}

The local error of an approximate evolution operator, defined as $\hat
{U}_{\mathrm{appr}}(\epsilon)-\hat{U}(\epsilon)$, is obtained by
comparing the Taylor expansion of $\hat{U}_{\mathrm{appr}}(\epsilon)$ with the
Taylor expansion of the exact evolution operator:%
\begin{equation}
\hat{U}(\epsilon)=e^{-i\epsilon\hat{H}}=1-i\epsilon\hat{H}%
-\frac{1}{2!}(\epsilon\hat{H})^{2}+\frac{i}{3!}(\epsilon\hat{H})^{3}%
+\mathcal{O}(\epsilon^{4}).\label{eq:exact_evol_op_taylor_expansion}%
\end{equation}
If the local error is $\mathcal{O(\epsilon}^{n+1})$, the method is said to be
of order $n$ because the global error for a finite time $t=N\epsilon$
is $\mathcal{O}(\epsilon^{n})$.

For the explicit Euler method, the Taylor expansion is identical to the
evolution operator (\ref{eq:exp_euler}) itself, and therefore the leading
order local error is $(\epsilon\hat{H})^{2}/2$. The Taylor expansion of the
implicit Euler method~(\ref{eq:imp_euler}) is the Neumann
series\cite{book_Stewart:1998}
\begin{equation}
\hat{U}_{\mathrm{impl}}(\epsilon)=(1+i\epsilon\hat{H})^{-1}=1-i\epsilon\hat
{H}+(i\epsilon\hat{H})^{2}-(i\epsilon\hat{H})^{3}+\mathcal{O}(\epsilon^{4});
\label{eq:imp_eu_taylor_expansion}%
\end{equation}
the leading order local error is $-(\epsilon\hat{H})^{2}/2$.

The Taylor expansions of the trapezoidal rule and implicit midpoint method are
obtained by composing Eqs.~(\ref{eq:exp_euler}) and
(\ref{eq:imp_eu_taylor_expansion}) with time steps $\epsilon/2$:
\begin{equation}
\hat{U}_{\mathrm{trap}}(\epsilon)=\hat{U}_{\mathrm{midp}}(\epsilon
)=1-i\epsilon\hat{H}-\frac{1}{2}(\epsilon\hat{H})^{2}+\frac{i}{4}(\epsilon
\hat{H})^{3}+\mathcal{O}(\epsilon^{4});
\label{eq:trap_rule_imp_mid_taylor_expansion}%
\end{equation}
the leading order local error of both methods is $i(\epsilon\hat{H})^{3}/12$.

Finally, the local error of the second-order differencing method is
\begin{equation}
-\frac{i}{3}(\epsilon\hat{H})^{3}+\mathcal{O}(\epsilon^{4}),
\label{eq:sod_error}%
\end{equation}
which is found by Taylor expanding $\psi_{\mathrm{sod}}(t-\epsilon)$, assumed
to be exact, in Eq.~(\ref{eq:sod_method}) to obtain
\begin{equation}
\psi_{\mathrm{sod}}(t+\epsilon)=\left(  1-i\epsilon\hat{H}-\frac{1}%
{2!}(\epsilon\hat{H})^{2}-\frac{i}{3!}(\epsilon\hat{H})^{3}+\mathcal{O}%
(\epsilon^{4})\right)  \psi_{\mathrm{sod}}(t)
\label{eq:sod_t_minus_dt_expanded}%
\end{equation}
and
\begin{equation}
\hat{U}_{\mathrm{sod}}=1-i\epsilon\hat{H}-\frac{1}{2!}(\epsilon\hat{H}%
)^{2}-\frac{i}{3!}(\epsilon\hat{H})^{3}+\mathcal{O}(\epsilon^{4}).
\label{eq:sod_prop_taylor_expansion}%
\end{equation}
Subtracting Eq.~(\ref{eq:exact_evol_op_taylor_expansion}) from
Eq.~(\ref{eq:sod_prop_taylor_expansion}) gives the local
error~(\ref{eq:sod_error}).

\subsection{\label{subsec:scalarprod}Unitarity}

Neither Euler method is unitary because
\begin{align}
\hat{U}_{\mathrm{expl}}(\epsilon)^{\dagger}\hat{U}_{\mathrm{expl}}(\epsilon)
&  =(1+i\epsilon\hat{H})(1-i\epsilon\hat{H})\nonumber\\
&  =1+\epsilon^{2}\hat{H}^{2} \label{eq:exp_eu_non_unitary}%
\end{align}
and
\begin{align}
\hat{U}_{\mathrm{impl}}(\epsilon)^{\dagger}\hat{U}_{\mathrm{impl}}(\epsilon)
&  =(1-i\epsilon\hat{H})^{-1}(1+i\epsilon\hat{H})^{-1}\nonumber\\
&  =\left(  (1+i\epsilon\hat{H})(1-i\epsilon\hat{H})\right)  ^{-1}\nonumber\\
&  =(1+\epsilon^{2}\hat{H}^{2})^{-1}\nonumber\\
&  =1-\epsilon^{2}\hat{H}^{2}+\mathcal{O}(\epsilon^{4}).
\label{eq:imp_eu_non_unitary}%
\end{align}

In contrast, both the trapezoidal rule and implicit midpoint methods are
unitary because
\begin{align}
\hat{U}_{\mathrm{trap}}(\epsilon)^{\dagger}\hat{U}_{\mathrm{trap}}(\epsilon)
&  =\left(  1+\frac{i\epsilon}{2}\hat{H}\right)  \left(  1-\frac{i\epsilon}%
{2}\hat{H}\right)  ^{-1}\left(  1+\frac{i\epsilon}{2}\hat{H}\right)
^{-1}\left(  1-\frac{i\epsilon}{2}\hat{H}\right) \nonumber\\
&  =\left(  1+\frac{i\epsilon}{2}\hat{H}\right)  \left(  1+\frac{i\epsilon}%
{2}\hat{H}\right)  ^{-1}\left(  1-\frac{i\epsilon}{2}\hat{H}\right)
^{-1}\left(  1-\frac{i\epsilon}{2}\hat{H}\right) \nonumber\\
&  =1\cdot1=1, \label{eq:trap_rule_unitary}%
\end{align}
(Proposition 1 was used in the first and Proposition 2 in the second line) and
because
\begin{align}
\hat{U}_{\mathrm{midp}}(\epsilon)^{\dagger}\hat{U}_{\mathrm{midp}}(\epsilon)
&  =\left(  1-\frac{i\epsilon}{2}\hat{H}\right)  ^{-1}\left(  1+\frac
{i\epsilon}{2}\hat{H}\right)  \left(  1-\frac{i\epsilon}{2}\hat{H}\right)
\left(  1+\frac{i\epsilon}{2}\hat{H}\right)  ^{-1}\nonumber\\
&  =\left(  1-\frac{i\epsilon}{2}\hat{H}\right)  ^{-1}\left(  1-\frac
{i\epsilon}{2}\hat{H}\right)  \left(  1+i\frac{\epsilon}{2}\hat{H}\right)
\left(  1+\frac{i\epsilon}{2}\hat{H}\right)  ^{-1}\nonumber\\
&  =1\cdot1=1 \label{eq:imp_midpoint_unitary}%
\end{align}
(Proposition 1 was used in the first line).

The analysis of its geometric properties is simplified if the second-order
differencing is represented by a $2\times2$ propagation matrix
\begin{equation}
\hat{U}_{\mathrm{sod}}(\epsilon):=%
\begin{pmatrix}
1-(2\epsilon\hat{H})^{2}, & -2i\epsilon\hat{H}\\
-2i\epsilon\hat{H}, & 1
\end{pmatrix}
\label{eq:sod_prop_matrix}%
\end{equation}
acting on a vector of $\psi$ at two different
times:\cite{Leforestier_Kosloff:1991}
\begin{equation}%
\begin{pmatrix}
\psi_{\mathrm{sod}}(t+\epsilon)\\
\psi_{\mathrm{sod}}(t)
\end{pmatrix}
=\hat{U}_{\mathrm{sod}}(\epsilon)%
\begin{pmatrix}
\psi_{\mathrm{sod}}(t-\epsilon)\\
\psi_{\mathrm{sod}}(t-2\epsilon)
\end{pmatrix}
. \label{eq:sod_prop_with_matrix}%
\end{equation}
Comparing the Hermitian conjugate $\hat{U}_{\mathrm{sod}}(\epsilon)^{\dag}$ of
$\hat{U}_{\mathrm{sod}}(\epsilon)$ with its inverse,
\begin{equation}
\hat{U}_{\mathrm{sod}}(\epsilon)^{-1}=%
\begin{pmatrix}
1, & 2i\epsilon\hat{H}\\
2i\epsilon\hat{H}, & 1-(2\epsilon\hat{H})^{2}%
\end{pmatrix}
, \label{eq:sod_prop_matrix_inverse}%
\end{equation}
found using $\det\hat{U}_{\mathrm{sod}}(\epsilon)=1,$ shows that the
second-order differencing is not unitary.

When $\hat{U}(\epsilon)$ is not unitary, we can obtain the time dependence of
the norm from
\begin{equation}
\Vert\psi(t+\epsilon)\Vert^{2}=\langle\psi(t)|\hat{U}(\epsilon)^{\dagger}%
\hat{U}(\epsilon)|\psi(t)\rangle. \label{eq:norm_t_p_dt}%
\end{equation}
For the explicit and implicit Euler methods, we find that%
\begin{align}
\Vert\psi_{\mathrm{expl}}(t+\epsilon)\Vert^{2}  &  =\Vert\psi_{\mathrm{expl}%
}(t)\Vert^{2}+\epsilon^{2}\langle\hat{H}^{2}\rangle_{\psi_{\mathrm{expl}}%
(t)},\label{eq:exp_eu_norm}\\
\Vert\psi_{\mathrm{impl}}(t+\epsilon)\Vert^{2}  &  =\Vert\psi_{\mathrm{impl}%
}(t)\Vert^{2}-\epsilon^{2}\langle\hat{H}^{2}\rangle_{\psi_{\mathrm{impl}}%
(t)}+\mathcal{O}(\epsilon^{3}), \label{eq:imp_eu_norm}%
\end{align}
where $\langle\hat{A}\rangle_{\psi}:=\langle\psi|\hat{A}|\psi\rangle$ denotes
the expectation value of operator $\hat{A}$ in state $\psi$.

Although the second-order differencing is not unitary, a conserved quantity
analogous to the inner product exists:

\textbf{Proposition 4.} The second-order differencing conserves the quantity%
\begin{equation}
(\langle\psi_{\mathrm{sod}}(t)|\phi_{\mathrm{sod}}(t-\epsilon)\rangle
+\langle\psi_{\mathrm{sod}}(t-\epsilon)|\phi_{\mathrm{sod}}(t)\rangle)/2.
\label{eq:sod_inner_prod}%
\end{equation}

The proof starts by projecting $\langle\phi_{\mathrm{sod}}(t)|$ on
Eq.~(\ref{eq:sod_method}), which yields
\begin{equation}
\langle\phi_{\mathrm{sod}}(t)|\psi_{\mathrm{sod}}(t+\epsilon)\rangle
=\langle\phi_{\mathrm{sod}}(t)|\psi_{\mathrm{sod}}(t-\epsilon)\rangle
-2i\epsilon\langle\phi_{\mathrm{sod}}(t)|\hat{H}|\psi_{\mathrm{sod}}%
(t)\rangle\text{.} \label{eq:sod_symplectic_intermediate_c}%
\end{equation}
Adding the complex conjugate of Eq.~(\ref{eq:sod_symplectic_intermediate_c})
to the analogue of Eq.~(\ref{eq:sod_symplectic_intermediate_c}), in which
$\psi$ and $\phi$ are exchanged, gives
\begin{align}
\langle\psi_{\mathrm{sod}}(t)|\phi_{\mathrm{sod}}(t+\epsilon)\rangle
+\langle\psi_{\mathrm{sod}}(t+\epsilon)|  &  \phi_{\mathrm{sod}}%
(t)\rangle\nonumber\\
&  =\langle\psi_{\mathrm{sod}}(t)|\phi_{\mathrm{sod}}(t-\epsilon
)\rangle+\langle\psi_{\mathrm{sod}}(t-\epsilon)|\phi_{\mathrm{sod}}%
(t)\rangle,\nonumber
\end{align}
completing the proof. As an immediate corollary, obtained by taking $\phi
=\psi$, the second-order differencing conserves the quantity $\mathrm{Re}%
\langle\psi_{\mathrm{sod}}(t)|\psi_{\mathrm{sod}}(t-\epsilon)\rangle$, which
is an analogue of the norm.\cite{Kosloff_Kosloff:1983}

\textbf{Proposition 5.}
The global error of the inner product between two quantum states
propagated by the second-order differencing is fourth-order in the time step,
i.e., $\langle\psi_{\mathrm{sod}}(t_{f})|\phi_{\mathrm{sod}}(t_{f})\rangle-\langle\psi(0)|\phi(0)\rangle=\mathcal{O}(\epsilon^{4}).$

Assuming that the wavepackets at $t=-\epsilon$ are known exactly, Proposition 4 implies
\begin{equation}
\langle\psi_{\mathrm{sod}}(t_{f}+\epsilon)|\phi_{\mathrm{sod}}(t_{f})\rangle+\langle\psi_{\mathrm{sod}}(t_{f})|\phi_{\mathrm{sod}}(t_{f}+\epsilon)\rangle=\langle\psi(0)|\phi(-\epsilon)\rangle+\langle\psi
(-\epsilon)|\phi(0)\rangle.\label{eq:sod_global_error_inner_prod}\end{equation}
By Taylor expanding
$\psi(-\epsilon)$ and $\phi(-\epsilon)$, and using
Eq.~(\ref{eq:sod_t_minus_dt_expanded}), we obtain\[
\langle\psi_{\mathrm{sod}}(t_{f})|\phi_{\mathrm{sod}}(t_{f})\rangle
-\frac{\epsilon^{2}}{2}\langle\psi_{\mathrm{sod}}(t_{f})|\hat{H}^{2}|\phi_{\mathrm{sod}}(t_{f})\rangle=\langle\psi(0)|\phi(0)\rangle
-\frac{\epsilon^{2}}{2}\langle\psi(0)|\hat{H}^{2}|\phi(0)\rangle
+\mathcal{O}(\epsilon^{4}).
\]
Rearranging the two sides gives
\begin{equation}
\langle\psi_{\mathrm{sod}}(t_{f})|\phi_{\mathrm{sod}}(t_{f})\rangle
-\langle\psi(0)|\phi(0)\rangle=\frac{\epsilon^{2}}{2}\left[  \langle
\psi_{\mathrm{sod}}(t_{f})|\hat{H}^{2}|\phi_{\mathrm{sod}}(t_{f})\rangle-\langle\psi(0)|\hat{H}^{2}|\phi(0)\rangle\right]  +\mathcal{O}(\epsilon^{4}).\label{eq:sod_global_error_innerprod_intermediate}\end{equation}
The global error of the second-order differencing method is second-order in
the time step and, therefore,
\begin{align}
\psi_{\mathrm{sod}}(t_{f})  & =\psi(t_{f})+\mathcal{O}(\epsilon^{2}),\label{eq:sod_global_error}\\
\phi_{\mathrm{sod}}(t_{f})  & =\phi(t_{f})+\mathcal{O}(\epsilon^{2}).\nonumber
\end{align}
Noting that under exact evolution, $\langle\psi(t_{f})|\hat{H}^{2}|\phi
(t_{f})\rangle=\langle\psi(0)|\hat{H}^{2}|\phi(0)\rangle$, we obtain
Proposition 5 by substituting Eq.~(\ref{eq:sod_global_error}) into
Eq.~(\ref{eq:sod_global_error_innerprod_intermediate}). 

\subsection{\label{subsec:symplecticity}Symplecticity}

Using a shorthand notation $\omega_{\text{appr}}|_{t}:=\omega(\psi
_{\text{appr}}(t),\phi_{\text{appr}}(t))$ and expressions $\hat{U}%
_{\text{appr}}(\epsilon)^{\dagger}\hat{U}_{\text{appr}}(\epsilon)$ from
Appendix~\ref{subsec:scalarprod} for the two Euler methods gives
\begin{align}
\omega_{\text{expl}}|_{t+\epsilon}  &  =\omega_{\text{expl}}|_{t}%
-2\hbar\epsilon^{2}\mathrm{Im}\langle\psi_{\mathrm{expl}}(t)|\hat{H}^{2}%
|\phi_{\mathrm{expl}}(t)\rangle\label{eq:exp_eu_symplectic}\\
\omega_{\text{impl}}|_{t+\epsilon}  &  =\omega_{\text{impl}}|_{t}%
+2\hbar\epsilon^{2}\mathrm{Im}\langle\psi_{\mathrm{impl}}(t)|\hat{H}^{2}%
|\phi_{\mathrm{impl}}(t)\rangle+\mathcal{O}(\epsilon^{3}),
\label{eq:imp_eu_symplectic}%
\end{align}
showing that neither first-order method is symplectic. In contrast, both the
trapezoidal rule and implicit midpoint methods are symplectic because they are unitary.

Finally, the second-order differencing is strictly not symplectic, but
Proposition 4 implies that the quantity
\begin{equation}
-\hbar\mathrm{Im}[\langle\psi_{\mathrm{sod}}(t)|\phi_{\mathrm{sod}}%
(t+\epsilon)\rangle+\langle\psi_{\mathrm{sod}}(t+\epsilon)|\phi_{\mathrm{sod}%
}(t)\rangle], \label{eq:sod_symplectic}%
\end{equation}
analogous to the symplectic two-form, is conserved. In fact,
Proposition 5 shows that the global error of the symplectic two-form is
$\mathcal{O}(\epsilon^{4})$.

\subsection{\label{subsec:comm}Commutation of the evolution operator with the
Hamiltonian}

Evolution operators of both Euler methods commute with the Hamiltonian:
\begin{align}
\lbrack\hat{H},\hat{U}_{\mathrm{expl}}(\epsilon)]  &  =[\hat{H},1-i\epsilon
\hat{H}]=0,\label{eq:exp_eu_comm}\\
\lbrack\hat{H},\hat{U}_{\mathrm{impl}}(\epsilon)]  &  =[\hat{H},(1+i\epsilon
\hat{H})^{-1}]=0, \label{eq:imp_eu_comm}%
\end{align}
where in Eq.~(\ref{eq:imp_eu_comm}), Proposition 2 was used. Applying
Proposition 3 to $\hat{A}=\hat{U}_{\mathrm{expl}}(\epsilon/2)$ and $\hat
{B}=\hat{U}_{\mathrm{impl}}(\epsilon/2)$ (or vice versa) then shows that the
evolution operators of both the trapezoidal rule and implicit midpoint methods
commute with the Hamiltonian. As for the second-order differencing, all
entries in $\hat{U}_{\mathrm{sod}}$ are polynomials in $\hat{H}$ and hence
commute with $\hat{H}$; as a result, $[\hat{H},\hat{U}_{\mathrm{sod}}]=0$ as well.

\subsection{\label{subsec:Econs}Energy conservation}

Neither Euler method is unitary and hence neither conserves the energy. In
contrast, both the trapezoidal rule and implicit midpoint methods conserve
energy because their evolution operators are unitary and commute with the Hamiltonian.

The second-order differencing does not conserve energy exactly but a conserved
quantity analogous to the energy has been
defined:\cite{Leforestier_Kosloff:1991} Applying $\langle\psi_{\mathrm{sod}%
}(t)|\hat{H}$ to Eq.~(\ref{eq:sod_method}) gives
\begin{equation}
\langle\psi_{\mathrm{sod}}(t)|\hat{H}|\psi_{\mathrm{sod}}(t+\epsilon
)\rangle=-2i\epsilon\langle\hat{H}^{2}\rangle_{\psi_{\mathrm{sod}}(t)}%
+\langle\psi_{\mathrm{sod}}(t)|\hat{H}|\psi_{\mathrm{sod}}(t-\epsilon)\rangle.
\label{eq:sod_energy_intermediate}%
\end{equation}
Because $\langle\hat{H}^{2}\rangle_{\psi_{\mathrm{sod}}(t)}$ is real, taking
the real part of Eq.~(\ref{eq:sod_energy_intermediate}) shows that
\begin{equation}
\mathrm{Re}\langle\psi_{\mathrm{sod}}(t)|\hat{H}|\psi_{\mathrm{sod}%
}(t+\epsilon)\rangle\label{eq:sod_energy_a}%
\end{equation}
is conserved.

\textbf{Proposition 6.}
The global error of the expectation value of energy of the quantum state
propagated by the second-order differencing is fourth-order in the time step,
i.e., $\langle\hat{H}\rangle_{\psi_{\mathrm{sod}}(t_{f})}-\langle\hat
{H}\rangle_{\psi(0)}=\mathcal{O}(\epsilon^{4})$.

From Eq.~(\ref{eq:sod_energy_a}),
\begin{equation}
\mathrm{Re}\langle\psi_{\mathrm{sod}}(t_{f})|\hat{H}|\psi_{\text{sod}}(t_{f}+\epsilon)\rangle=\mathrm{Re}\langle\psi(-\epsilon)|\hat{H}|\psi(0)\rangle.
\end{equation}
Assuming, as in the proof of Proposition 5, that $\psi(-\epsilon)$ is known
exactly, by Taylor expanding $\psi(-\epsilon)$ and using
Eq.~(\ref{eq:sod_t_minus_dt_expanded}), we obtain
\begin{equation}
\langle\hat{H}\rangle_{\psi_{\mathrm{sod}}(t_{f})}-\langle\hat{H}\rangle
_{\psi(0)}=\frac{\epsilon^{2}}{2}\left[  \langle\hat{H}^{3}\rangle
_{\psi_{\mathrm{sod}}(t_{f})}-\langle\hat{H}^{3}\rangle_{\psi(0)}\right]
+\mathcal{O}(\epsilon^{4}).\label{eq:global_error_sod_energy_intermediate}\end{equation}
Invoking Eq.~(\ref{eq:sod_global_error}) and identity $\langle\hat{H}^{3}\rangle_{\psi(t_{f})}=\langle\hat{H}^{3}\rangle_{\psi(0)}$ completes the
proof of Proposition 6.

\subsection{\label{subsec:symm}Symmetry}

\textbf{Proposition 7.} The adjoint of an evolution operator has the following
properties:%
\begin{align}
(\hat{U}(\epsilon)^{\ast})^{\ast}  &  =\hat{U}(\epsilon),\label{eq:adj_adj}\\
(\hat{U}_{1}(\epsilon)\hat{U}_{2}(\epsilon))^{\ast}  &  =\hat{U}_{2}%
(\epsilon)^{\ast}\hat{U}_{1}(\epsilon)^{\ast},\label{eq:adj_prod}\\
&  \hat{U}(\epsilon)\hat{U}(\epsilon)^{\ast}\text{ is symmetric.}
\label{eq:sym_op_w_adj}%
\end{align}

The first and second properties mean, respectively, that the adjoint operation
$^{\ast}$ is an involution and an antiautomorphism on the group of invertible
operators, while the last property provides the simplest recipe for
constructing a symmetric method---by composing a general method with its
adjoint, with both composition coefficients of $1/2$. All three properties
follow directly from the definition: the first because $(\hat{U}%
(\epsilon)^{\ast})^{\ast}=(\hat{U}(-\epsilon)^{\ast})^{-1}=(\hat{U}%
(\epsilon)^{-1})^{-1},$ the second because $(\hat{U}_{1}(\epsilon)\hat{U}%
_{2}(\epsilon))^{\ast}=(\hat{U}_{1}(-\epsilon)\hat{U}_{2}(-\epsilon
))^{-1}=\hat{U}_{2}(-\epsilon)^{-1}\hat{U}_{1}(-\epsilon)^{-1}$, and the third
by applying Eq.~(\ref{eq:adj_prod}) to the product of $\hat{U}$ and $\hat
{U}^{\ast}$, and using Eq.~(\ref{eq:adj_adj}).

The explicit and implicit Euler methods are adjoints of each other, which
follows from
\begin{equation}
\hat{U}_{\mathrm{expl}}(-\epsilon)^{-1}=(1+i\hat{H}\epsilon)^{-1}=\hat
{U}_{\mathrm{impl}}(\epsilon) \label{eq:exp_imp_eu_adjoint}%
\end{equation}
and Eq.~(\ref{eq:adj_adj}). Therefore, neither method is symmetric. In
contrast, the trapezoidal rule and implicit midpoint methods are both
symmetric, which follows from Eq.~(\ref{eq:sym_op_w_adj}) applied to the
composition of the explicit and implicit Euler methods with composition
coefficients $1/2$.

Taking the inverse of $\hat{U}_{\mathrm{sod}}(-\epsilon)$ gives
\begin{equation}
\hat{U}_{\mathrm{sod}}(-\epsilon)^{-1}=%
\begin{pmatrix}
1, & -2i\epsilon\hat{H}\\
-2i\epsilon\hat{H}, & 1-(2\epsilon\hat{H})^{2}%
\end{pmatrix}
=%
\begin{pmatrix}
0 & 1\\
1 & 0
\end{pmatrix}
\hat{U}_{\mathrm{sod}}(\epsilon)%
\begin{pmatrix}
0 & 1\\
1 & 0
\end{pmatrix}
, \label{eq:sod_prop_matrix_inverse_a}%
\end{equation}
implying that the second order differencing is symmetric if the sequence of
wavefunctions is reversed when taking the inverse.

\subsection{\label{subsec:trev}Time reversibility}

For an elementary time step $\epsilon$, time reversibility is a direct
consequence of the symmetry of the operator: if the operator is symmetric,
i.e., if $\hat{U}(-\epsilon)^{-1}=\hat{U}(\epsilon)$, then a forward
propagation is exactly cancelled by the immediately following backward
propagation:
\begin{equation}
\hat{U}(-\epsilon)\hat{U}(\epsilon)=\hat{U}(-\epsilon)\hat{U}(-\epsilon
)^{-1}=1. \label{eq:t_rev}%
\end{equation}
This argument is easily extended, by induction, to a forward propagation for
$N$ steps followed by a backward propagation for $N$ steps:%
\[
\hat{U}(-\epsilon)^{N}\hat{U}(\epsilon)^{N}=1.
\]
As a result, the Euler methods are not time-reversible, whereas the
trapezoidal rule, implicit midpoint, and second-order differencing methods are.

\subsection{\label{subsec:stability}Stability}

The explicit Euler method is unstable because, using Eq.~(\ref{eq:exp_eu_norm}%
),
\begin{align}
\Vert\psi(t+\epsilon)-\phi(t+\epsilon)\Vert^{2}  &  =\Vert\psi(t)-\phi
(t)\Vert^{2}+\epsilon^{2}\langle\hat{H}^{2}\rangle_{\psi(t)-\phi
(t)}\nonumber\\
&  \geq(1+\epsilon^{2}E_{\text{min}}^{2})\Vert\psi(t)-\phi(t)\Vert^{2},
\label{exp_eu_l2norm_difference}%
\end{align}
as long as $\hat{H}$ has no eigenvalue in the finite interval $(-E_{\text{min}%
},E_{\text{min}})$; composing the above inequality $N$ times shows that%
\begin{equation}
\Vert\psi(N\epsilon)-\phi(N\epsilon)\Vert^{2}\geq(1+\epsilon^{2}E_{\text{min}%
}^{2})^{N}\Vert\psi(0)-\phi(0)\Vert^{2}\rightarrow\infty
\label{eq:exp_eu_unstable}%
\end{equation}
as $N\rightarrow\infty$ for $\psi(0)\neq\phi(0)$.

Asymptotic stability of the implicit Euler method follows, using
Eq.~(\ref{eq:imp_eu_norm}), from an analogous inequality%
\begin{align}
\Vert\psi(t)-\phi(t)\Vert^{2}  &  =\Vert\psi(t+\epsilon)-\phi(t+\epsilon
)\Vert^{2}+\epsilon^{2}\langle\hat{H}^{2}\rangle_{\psi(t+\epsilon
)-\phi(t+\epsilon)}\nonumber\\
&  \geq(1+\epsilon^{2}E_{\text{min}}^{2})\Vert\psi(t+\epsilon)-\phi
(t+\epsilon)\Vert^{2}, \label{eq:imp_eu_asym_stable_intermediate}%
\end{align}
which implies
\begin{equation}
\Vert\psi(N\epsilon)-\phi(N\epsilon)\Vert^{2}\leq(1+\epsilon^{2}E_{\text{min}%
}^{2})^{-N}\Vert\psi(0)-\phi(0)\Vert^{2}\rightarrow0
\label{eq:imp_eu_asym_stable}%
\end{equation}
as $N\rightarrow\infty$.

Both the trapezoidal rule and implicit midpoint methods are unitary, and
therefore
\begin{equation}
\Vert\psi(t+\epsilon)-\phi(t+\epsilon)\Vert=\Vert\psi(t)-\phi(t)\Vert;
\label{eq:trap_rule_imp_mid_stable}%
\end{equation}
as a result, both methods are stable but not asymptotically stable.

Following Leforestier \textit{et al.}\cite{Leforestier_Kosloff:1991} and
slightly abusing notation, the stability of the second-order differencing is
analyzed by examining the eigenvalues
\begin{equation}
\lambda_{1,2}=1-2\epsilon^{2}\hat{H}^{2}\pm2\epsilon\hat{H}(\epsilon^{2}%
\hat{H}^{2}-1)^{\frac{1}{2}} \label{eq:sod_eigenvalues}%
\end{equation}
of $\hat{U}_{\mathrm{sod}}(\epsilon)$. For the method to be stable, the
eigenvalues must be complex units (i.e., $\left\vert \lambda_{1,2}\right\vert
=1$), which is equivalent to requiring
\begin{equation}
\epsilon^{2}\hat{H}^{2}-1<0. \label{eq:sod_stability_condition}%
\end{equation}
Otherwise, the magnitude of one of the eigenvalues is greater than unity and
the method is unstable.\cite{Leforestier_Kosloff:1991} For the stability
criterion to be met, the condition~(\ref{eq:sod_stability_condition}) must be
satisfied for all energy eigenstates and, therefore, the method is stable only
for time steps\cite{Leforestier_Kosloff:1991}
\begin{equation}
\epsilon<\frac{1}{E_{\mathrm{max}}}, \label{eq:sod_critical_timestep}%
\end{equation}
where $E_{\mathrm{max}}$ is the largest eigenvalue of the Hamiltonian operator
approximated by a finite matrix.

\section{\label{appendixb}Exponential convergence with grid density}

\begin{figure}
[pth]\includegraphics[scale=1.0]{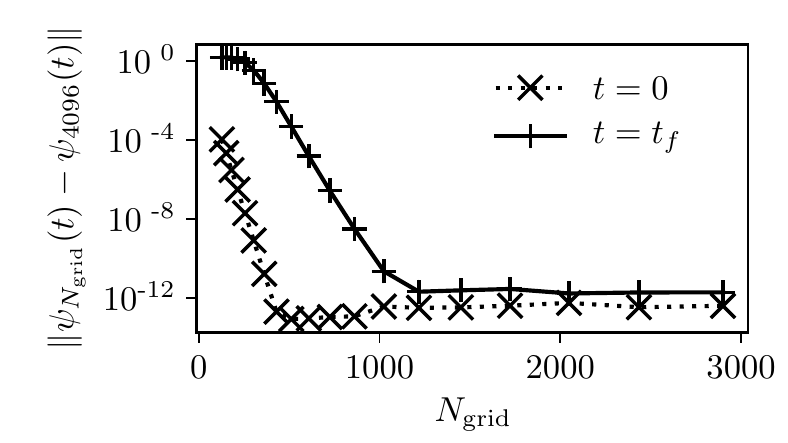}
\caption{Convergence of the wavepacket (measured by the $L^{2}$-norm of the error) at the initial ($t=0$) and final ($t=t_{f}$)
times with the increasing number of grid points. The optimally composed $10^{\mathrm{th}}$-order trapezoidal rule with $\Delta t=0.25$ a.u. was used.}\label{fig:gridconv}%

\end{figure}

Figure~\ref{fig:gridconv} demonstrates the exponential convergence of the
wavefunction with the increasing number of grid points. In order to have
balanced position and momentum grids, the ranges as well as the densities of
both the position and momentum grids were increased by a factor of $\sqrt{2}$
for every increase in the number of grid points by a factor of two. Comparison
of wavepackets on grids with different densities was carried out by
trigonometric interpolation of the wavepacket on the sparser grid.\newpage

\bibliographystyle{aipnum4-1}
\bibliography{integrators_nonadiabatic}

\end{document}